\documentclass[
  aps,
  prd,
  reprint,
  superscriptaddress,
  nofootinbib,
  amsmath,
  amssymb
]{revtex4}

\usepackage{bm}
\usepackage{mathrsfs}
\usepackage{dsfont}

\usepackage{graphicx}
\usepackage{booktabs}
\usepackage{pdflscape}
\usepackage{rotating}

\usepackage[dvipsnames,svgnames]{xcolor}
\usepackage[normalem]{ulem}

\colorlet{forestgreen}{ForestGreen}

\usepackage{comment}
\usepackage{csquotes}
\usepackage{url}

\usepackage{hyperref}

\hypersetup{
  colorlinks=true,
  linkcolor=blue,
  citecolor=violet,
  filecolor=violet,
  urlcolor=blue
}

\begin{document}

\title{Scalar quasinormal modes of Schwarzschild--anti-de Sitter black holes:\\
spectral analysis and generalized boundary conditions}

\author{Davide Batic}
\email{davide.batic@ku.ac.ae}
\affiliation{Mathematics Department, Khalifa University of Science and Technology, PO Box 127788, Abu Dhabi, United Arab Emirates}

\author{Alan S. Cornell}
\email{acornell@uj.ac.za}
\affiliation{Department of Physics, University of Johannesburg, PO Box 524, Auckland Park 2006, South Africa}

\author{Denys Dutykh}
\email{denys.dutykh@ku.ac.ae}
\affiliation{Mathematics Department, Khalifa University of Science and Technology, PO Box 127788, Abu Dhabi, United Arab Emirates}

\date{\today}

\begin{abstract}

We study quasinormal modes (QNMs) of a minimally coupled massless scalar field on four-dimensional Schwarzschild--anti-de Sitter black holes using a Chebyshev spectral method. After compactifying the exterior domain, the radial problem is formulated as a quadratic matrix pencil in the dimensionless frequency. For the standard Dirichlet, or vanishing-field, boundary condition at the conformal AdS boundary, we reproduce the known scalar spectra across small, intermediate, and large black holes, including long overtone sequences and the expected approach to pure-AdS normal modes in the small black hole limit. We then deform the AdS boundary condition by imposing a generalized relation between the two independent asymptotic coefficients of the massless scalar. This deformation is treated as a generalized coefficient boundary condition for the massless scalar, and not as the usual alternative quantization for scalars in the Breitenlohner-Freedman window. The Dirichlet endpoint recovers the stable standard spectrum. For every non-Dirichlet value examined, and for representative small, intermediate, and large black holes, we find an additional mode with positive imaginary part, signaling a boundary-condition-induced instability. A near-Dirichlet refinement finds no finite critical angle down to the smallest deformation probed.
\end{abstract}

\maketitle


\section{Introduction}\label{sec:intro}

Quasinormal modes (QNMs) are the characteristic damped oscillations of black holes. They encode how perturbations relax after a disturbance and provide a coordinate-invariant description of the dissipative part of black hole dynamics. In asymptotically flat spacetimes, the standard QNM problem is defined by imposing purely ingoing behavior at the event horizon and purely outgoing behavior at spatial infinity. In asymptotically de Sitter spacetimes, the role of spatial infinity is replaced by the cosmological horizon, where one again imposes an outgoing radiative condition. In both cases, the boundary conditions are radiative at both ends of the physical domain, and the resulting eigenvalue problem is intrinsically non-self-adjoint, leading to a discrete set of complex frequencies. Comprehensive reviews of the role of QNMs in black hole physics can be found in Refs \cite{Kokkotas1999LR,Berti2009CQG,Konoplya2011}. 

The situation is qualitatively different in asymptotically anti-de Sitter (AdS) spacetimes. The conformal boundary of AdS spacetimes is timelike, and hence, the large radius endpoint is not a wave zone. There is no canonical outgoing wave condition at infinity. Instead, one must prescribe boundary data at the AdS boundary. Thus, in AdS spacetimes, the choice of boundary condition is not a technical detail added after the differential equation has been written down. It is part of the definition of the spectral problem itself. This feature is central both mathematically and physically. On the mathematical side, AdS is not globally hyperbolic, and the evolution problem is well posed only after an admissible boundary condition has been specified at the conformal boundary. The self-adjoint extension analysis of Refs \cite{IshibashiWald2003,IshibashiWald2004} provides an important framework for classifying admissible boundary conditions at the timelike conformal boundary of AdS spacetime. Their construction addresses the well-posedness of the underlying evolution problem by requiring the spatial part of the wave operator to admit an appropriate positive self-adjoint extension, with different extensions corresponding to different allowed boundary conditions at infinity. A rigorous framework for asymptotically AdS QNMs with general boundary conditions was developed by Warnick, who formulated the modes as eigenvalues associated with the generator
of the solution semigroup and proved discreteness properties of the resulting spectrum under broad assumptions \cite{Warnick2015}.

On the physical side, in the context of gauge/gravity duality, AdS QNM frequencies determine relaxation timescales of perturbations in the dual field theory, so the boundary condition at infinity fixes which source-response problem is being studied \cite{HorowitzPRD2000,Berti2009CQG}. The black-hole QNM problem is, however, not itself a self-adjoint eigenvalue problem. In addition to the boundary condition imposed at conformal infinity, one requires purely ingoing behavior at the event horizon, thereby allowing perturbative flux to be absorbed by the black hole. This radiative horizon condition renders the complete QNM boundary-value problem non-Hermitian and gives rise to a discrete spectrum of complex frequencies. Accordingly, the Ishibashi-Wald analysis should be understood as constraining the admissible asymptotic data at the AdS boundary, rather than as furnishing a self-adjoint formulation of the full QNM problem. In this sense, the AdS boundary condition fixes the domain of the problem at infinity, while the ingoing horizon condition encodes dissipation through the black hole.

The first systematic numerical study of scalar QNMs of Schwarzschild-AdS (SAdS) black holes was carried out by Ref. \cite{HorowitzPRD2000}, who imposed purely ingoing behavior at the event horizon and a vanishing field condition at the AdS boundary. Their power series method became the standard benchmark for scalar perturbations in SAdS. Subsequent studies extended the analysis to other fields, different black hole sizes, and higher overtones. Ref. \cite{Konoplya2002} studied the small black hole regime and showed that the scalar modes approach the normal modes of pure AdS as the horizon radius tends to zero, whilst Ref. \cite{CardosoKonoplyaLemos2003} investigated the scalar, electromagnetic and gravitational spectra in detail, including the asymptotic large overtone structure. Further numerical checks and alternative algorithms include the asymptotic iteration method \cite{ChoCornellDoukasNaylor2010}, the interpolation matrix method  \cite{LinCQG2017}, and the continued fraction calculation  \cite{DaghighGreenMorey2023}.

Boundary conditions in AdS spaces become still subtler for fields beyond the minimally coupled scalar. Related Robin-type boundary conditions and their dynamical consequences have been studied in different scalar settings, including conformally coupled scalar dynamics on SAdS backgrounds \cite{FicekMaliborski2024}. For Maxwell perturbations, a field-vanishing condition is reflective, but it is not the most general condition associated with vanishing energy flux at the AdS boundary. Imposing vanishing energy flux leads to two possible Robin branches, one reproducing the usual spectrum found in earlier calculations and another giving an additional branch of electromagnetic QNMs \cite{CardosoPRD2001,WangHerdeiroSampaio2015}. For gravitational perturbations, the choice of AdS-boundary condition is also sector dependent. Moss and Norman discussed gravitational QNMs of AdS black holes and the role of AdS/CFT-motivated boundary conditions \cite{MossNorman2002}. In the scalar-type, or polar, gravitational sector of global AdS$_4$--Schwarzschild, Michalogiorgakis and Pufu showed that fixing the boundary metric leads naturally to a Robin condition for the corresponding master field \cite{MichalogiorgakisPufu2007}. This should be distinguished from the older practice of imposing field-vanishing conditions on the gravitational master variables. These examples show that the QNM spectrum of an AdS black hole is not a single object until the boundary condition at infinity has been specified.

The present work focuses on the scalar sector of four-dimensional Schwarzschild-AdS black holes. Our first aim is to construct and validate a high-precision Chebyshev spectral implementation for the standard minimally coupled massless scalar with the Dirichlet, or vanishing-field, condition at the conformal AdS boundary. This problem provides a stringent test because a large body of independent numerical data is available. We formulate the radial equation as a quadratic matrix pencil in the dimensionless frequency $\Omega=\omega R$, compactify the exterior domain, factor out the asymptotic QNM behavior, and identify physical modes by matching spectra computed at three different spectral resolutions. We compare the resulting frequencies with the Horowitz-Hubeny power-series method, Konoplya's small black hole data, the Cardoso-Konoplya-Lemos overtone tables, the asymptotic iteration method, the Lin-Qian matrix method, and the continued fraction calculation of Daghigh, Green and Morey. Our second aim is to use the same spectral framework to probe how the scalar spectrum changes when the standard Dirichlet condition is deformed. For the minimally coupled massless scalar, the usual finite-energy prescription at the AdS boundary selects the Dirichlet branch. By contrast, a genuine Robin family in the standard alternative-quantization sense is naturally associated with scalar fields for which both asymptotic falloffs are admissible, such as fields in the Breitenlohner-Freedman window \cite{IshibashiWald2004,Kinoshita2024}. Here, we pursue a different question, namely we keep the same minimally coupled massless scalar field, but impose a generalized coefficient relation between the two independent asymptotic data in its large-radius expansion. The parameter controlling this relation interpolates between the standard Dirichlet endpoint and a coefficient-Neumann endpoint, with intermediate values defining a controlled deformation of the scalar QNM problem. This deformation should not be confused with the usual Robin boundary condition arising in alternative quantization. Rather, it should be viewed as a generalized source-response relation for the minimally coupled massless scalar. It allows us to ask whether the stability of the standard scalar SAdS spectrum is robust under non-Dirichlet deformations of the boundary data. Note that Robin-type AdS boundary conditions have been studied in related, but distinct, scalar settings. For example, Ref.~\cite{Kinoshita2024} analyzes a scalar with
two admissible falloffs on a planar AdS black hole, while Ref.~\cite{FicekMaliborski2024} studies conformally coupled scalar dynamics on
a SAdS background. These works provide useful context for the role of boundary conditions in AdS, but they do not represent the same minimally coupled massless scalar problem considered here. The main result of this second analysis is that the generalized coefficient condition generically produces an unstable mode. The Dirichlet endpoint is stable and reproduces the standard scalar spectrum, but for every nonzero value of the deformation parameter $\theta$ tested we find an additional mode with $\operatorname{Im}\Omega>0$. With our convention $e^{-i\omega t}$, this corresponds to exponential growth in time. The instability appears in the near-global-AdS regime, in the intermediate regime, and in the large black hole regime. A near-Dirichlet refinement shows that no finite critical angle is observed down to $\theta=\pi/64$. Within the numerical range probed, the instability appears as soon as the Dirichlet endpoint is deformed.

The paper is organized as follows. In Sec.~\ref{sec:geometry} we introduce the Schwarzschild-AdS geometry, fix conventions, and derive the scalar radial equation. We also summarize the standard boundary conditions used for scalar, electromagnetic and gravitational perturbations in the SAdS literature. Section~\ref{SecSpectralMethod} describes the Chebyshev spectral method, the quadratic matrix-pencil formulation, the resolution-triplet matching procedure, and the modification needed to impose the generalized coefficient boundary condition. In Sec.~\ref{sec:results}, we present the numerical spectra. We first benchmark the Dirichlet scalar problem against existing literature and then study the instability induced by the generalized coefficient condition. Finally, Sec.~\ref{sec:conclusions} summarizes the results and discusses possible extensions to electromagnetic and gravitational perturbations.


\section{Metric and equations of motion}\label{sec:geometry}

To introduce the QNM eigenvalue problem and fix our conventions, we first review the minimally coupled massless scalar equation on a four-dimensional SAdS background and then quote the analogous electromagnetic and gravitational master equations. We use units $c=G_N=\hbar=1$. The line element is written as \cite{HorowitzPRD2000,CardosoPRD2001}
\begin{equation}\label{LE}
ds^2=-f(r)dt^2+\frac{dr^2}{f(r)}+r^2\left(d\vartheta^2+\sin^2\vartheta\,d\varphi^2\right) \; ,\qquad f(r)=1-\frac{2M}{r}+\frac{r^2}{R^2} \; ,
\end{equation}
where $M$ is the black hole mass parameter and $R$ is the AdS radius, related to the negative cosmological constant by $\Lambda=-3/R^2$. It is convenient to replace $M$ by the event-horizon radius $r_+$. Since $r_+$ is the largest positive root of $f(r)$, the condition $f(r_+)=0$ gives
\begin{equation}\label{hor}
    2M=r_+\left(1+\frac{r_+^2}{R^2}\right) \; ,
\end{equation}
and hence
\begin{equation}\label{fr0}
    f(r)=1+\frac{r^2}{R^2}-\frac{r_+}{r}\left(1+\frac{r_+^2}{R^2}\right) \; .
\end{equation}
We define the tortoise coordinate by
\begin{equation}
\frac{dr_*}{dr}=\frac{1}{f(r)}\; ,\qquad
\frac{d}{dr_*}=f(r)\frac{d}{dr} \; .
\end{equation}
For a minimally coupled massless scalar field we use the separation
\begin{equation}\label{ansatzgs}
\phi(t,r,\vartheta,\varphi)=e^{-i\omega t}Y_{\ell m}(\vartheta,\varphi)\frac{\psi(r)}{r} \; .
\end{equation}
The scalar equation takes the Schr\"odinger-type form \cite{HorowitzPRD2000, CardosoPRD2001}
\begin{equation}\label{ODE01}
\frac{d^2\psi}{dr_*^2}+\left[\omega^2-U(r)\right]\psi(r)=0 \; ,
\end{equation}
or equivalently
\begin{equation}\label{ODE02}
    f(r)\frac{d}{dr}\left(f(r)\frac{d\psi}{dr}\right)
    +\left[\omega^2-U(r)\right]\psi(r)=0 \; ,
\end{equation}
where
\begin{equation}\label{Uepsilon}
U(r)=f(r)\left[\frac{1}{r}\frac{df}{dr}+\frac{\ell(\ell+1)}{r^2}
\right] \; .
\end{equation}
The allowed multipoles for scalar perturbations are $\ell=0,1,2,\ldots$. With the time dependence $e^{-i\omega t}$, decaying modes have $\textcolor{blue}{\operatorname{Im}}\,\omega<0$. At the event horizon, the QNM condition is purely ingoing behavior,
\begin{equation}
\psi(r)\sim e^{-i\omega r_*} \; ,\qquad r\to r_+ \; .
\end{equation}
The boundary condition at the AdS boundary depends on the field and on the physical problem under consideration (see Table~\ref{tab:ads_bc_spin} for a general overview). For the standard minimally coupled massless scalar benchmark one imposes
\begin{equation}
\psi(r)\sim r^{-2} \; ,\qquad r\to\infty \; .
\end{equation}
More general scalar boundary conditions will be specified separately. Finally, we introduce the dimensionless variables $x=r/R$, $x_+=r_+/R$, $\Omega=\omega R$, and $\mu=M/R$. From Eq.~\eqref{fr0}, one obtains
\begin{equation}\label{fx}
f(x)=1+x^2-\frac{x_+(1+x_+^2)}{x}
=(x-x_+)\frac{x^2+x_+x+x_+^2+1}{x} \; .
\end{equation}
Once $x_+$ is specified, the dimensionless mass parameter is recovered from
\begin{equation}\label{mu_relation}
2\mu=x_+(1+x_+^2) \; ,
\end{equation}
which is the dimensionless form of Eq.~\eqref{hor}. Thus, after measuring all lengths in units of the AdS radius $R$, the geometry is completely characterized by the single dimensionless parameter $x_+=r_+/R$. Fixing different values of $x_+$ is therefore equivalent to choosing different values of the dimensionless mass parameter $\mu$ according to Eq.~\eqref{mu_relation}. Since
\[
\frac{d\mu}{dx_+}=\frac{1+3x_+^2}{2}>0 \; ,
\]
the relation \eqref{mu_relation} is one-to-one for $x_+>0$. In what follows we label the spectra by $x_+$, which also allows a direct distinction between small $(x_+\ll 1)$, intermediate $(x_+\sim 1)$, and large $(x_+\gg 1)$ SAdS black holes. Finally, in the new variable $x$ the radial equation \eqref{ODE02} becomes
\begin{equation}\label{ODE03}
    f(x)\frac{d}{dx}\left(f(x)\frac{d\psi}{dx}\right)
    +\left[\Omega^2-U(x)\right]\psi(x)=0 \; ,\quad\Omega=\omega R \; ,
\end{equation}
where
\begin{equation}\label{Uepsilonx}
U(x)=f(x)\left[\frac{1}{x}\frac{df}{dx}+\frac{\ell(\ell+1)}{x^2}
\right] \; .
\end{equation}
\begin{table*}[t]
\caption{\textit{
Representative AdS-boundary conditions employed in Schwarzschild--AdS
quasinormal-mode calculations. We use the convention \(e^{-i\omega t}\) and define the tortoise coordinate by \(dr_*/dr=1/f(r)\). The event-horizon condition is purely ingoing in all cases; the main distinction lies in the condition imposed at the timelike conformal AdS boundary. The two scalar rows describe different physical settings. The first row is the minimally coupled massless scalar benchmark studied in the present work. The second row refers to scalar theories whose two AdS falloffs are both admissible, as in the Breitenlohner--Freedman window. The generalized coefficient boundary
condition introduced later for the minimally coupled massless scalar should therefore be viewed as a deformation of the massless scalar spectral problem, not as the standard Robin family associated with alternative quantization.
}}
\label{tab:ads_bc_spin}
\centering
\scriptsize
\renewcommand{\arraystretch}{1.35}
\setlength{\tabcolsep}{4pt}

\begin{tabular}{@{}p{0.055\textwidth}
                p{0.18\textwidth}
                p{0.25\textwidth}
                p{0.20\textwidth}
                p{0.25\textwidth}@{}}
\toprule
\textbf{Spin} &
\textbf{Field} &
\textbf{Representative setting} &
\textbf{Event-horizon condition} &
\textbf{AdS-boundary condition} \\
\midrule

\(s=0\) &
Minimally coupled massless scalar &
Standard scalar Schwarzschild--AdS QNM benchmark
\cite{HorowitzPRD2000,Konoplya2002,CardosoKonoplyaLemos2003,DaghighGreenMorey2023}. &
Purely ingoing:
\[
    \psi\sim e^{-i\omega r_*},
    \qquad r\to r_+ .
\]
&
Dirichlet, or normalizable, falloff:
\[
    \Phi\to0 .
\]
With \(\Phi=\psi/r\), this gives
\[
    \psi\sim r^{-2},
    \qquad r\to\infty .
\]
\\
\midrule

\(s=0\) &
Scalar fields with two admissible AdS falloffs &
Scalars in the Breitenlohner--Freedman or alternative-quantization window;
Ref.~\cite{Kinoshita2024} provides a related planar-AdS black-hole example. &
Purely ingoing, or regular after factoring out the ingoing horizon behaviour. &
Mixed/Robin condition relating the slow and fast falloff coefficients. For
example, writing \(\rho=R/r\),
\[
    \Phi\sim \rho^{\Delta_-}\phi_1
    +\rho^{\Delta_+}\phi_2,
\]
one may impose
\[
    \phi_2=\kappa\phi_1 .
\]
Dirichlet and Neumann arise as limiting or special cases. \\
\midrule

\(s=1\) &
Maxwell field &
Classical electromagnetic Schwarzschild--AdS calculations
\cite{CardosoPRD2001,CardosoKonoplyaLemos2003,DaghighGreenMorey2023}. &
Purely ingoing:
\[
    \psi\sim e^{-i\omega r_*}.
\]
&
Standard reflective field-vanishing condition:
\[
    \psi\to0,
    \qquad
    \psi\sim r^{-1}.
\]
If
\[
    \psi\sim a_0+\frac{a_1}{r}+\cdots,
\]
this corresponds to \(a_0=0\). \\
\midrule

\(s=1\) &
Maxwell field &
Vanishing-energy-flux formulation
\cite{WangHerdeiroSampaio2015,WangChenTongPanJing2021}. &
Purely ingoing. &
Vanishing energy flux allows two reflective branches. In a
Regge--Wheeler-type variable,
\[
    \psi\sim a_0+\frac{a_1}{r}+\cdots,
\]
the two branches may be written as
\[
    a_0=0
    \qquad \text{or} \qquad
    a_1=0 .
\]
In Teukolsky variables these correspond to two Robin boundary conditions. \\
\midrule

\(s=2\) &
Gravitational perturbations &
Early Regge--Wheeler/Zerilli Schwarzschild--AdS calculations
\cite{CardosoPRD2001,CardosoKonoplyaLemos2003}. &
Purely ingoing. &
Field-vanishing conditions for the gravitational master wavefunctions, as
adopted in several early numerical studies. \\
\midrule

\(s=2\) &
Gravitational perturbations &
AdS/CFT-motivated and fixed-boundary-metric formulations
\cite{MossNorman2002,MichalogiorgakisPufu2007}.&
Purely ingoing. &
The appropriate boundary condition depends on the gravitational sector and on the boundary prescription. Moss--Norman discuss AdS/CFT-motivated gravitational boundary conditions, while Michalogiorgakis--Pufu show that the polar (scalar-type) gravitational sector obeys a Robin-type condition when the
boundary metric is kept fixed. \\
\bottomrule
\end{tabular}
\end{table*}
Introducing the dimensionless radial coordinate $z=x/x_+$, which maps the event horizon to $z=1$, Eq.~\eqref{ODE03} takes the equivalent form
\begin{equation}\label{ourODE}
    f(z)\frac{d}{dz}\left(f(z)\frac{d\psi}{dz}\right)+\left[x_+^2\Omega^2-V(z)\right]\psi(z)=0 \;,\qquad
    V(z)=f(z)\left[\frac{1}{z}\frac{df}{dz}+\frac{\ell(\ell+1)}{z^2}\right] \;,\qquad
    \Omega=\omega R \;,
\end{equation}
with $f(z)$ given by
\begin{equation}\label{fz}
f(z)=(z-1)\frac{x^2_+ z^2+x^2_+ z+x^2_++1}{z} \;.
\end{equation}
In the following analysis, we focus on computing the QNMs for the spectral problem stated in Eq.~\eqref{ourODE}. We write $\Omega$ as $\Omega=\Omega_R+i\Omega_I$, where $\Omega_I<0$ corresponds to a perturbation that decays in time for the convention $e^{-i\omega t}$.

\subsubsection{The scalar case with Dirichlet boundary conditions at the conformal AdS boundary}

For the standard Dirichlet problem, the radial field is required to be purely ingoing at the event horizon and to satisfy the vanishing-field condition at the conformal AdS boundary. To apply the Chebyshev spectral method, we first determine the asymptotic behavior of the radial solution $\psi$ as $z\to1^+$ and $z\to+\infty$, factor out the corresponding QNM behavior, and then compactify the domain onto the finite interval $[-1,1]$, where the regular part of the solution is expanded in Chebyshev polynomials. We therefore analyze the radial field in the following two asymptotic regions
\begin{enumerate}
\item
{\textit{Asymptotic behavior as $z\to 1^+$}}: The point $z=1$ is a simple zero of $f(z)$ because $f(1)=0$ and $f'(1)=1+3x_+^2\neq0$, where the prime denotes differentiation with respect to $z$. It is convenient to write Eq.~\eqref{ourODE} in the form
\begin{equation}\label{ODEZ}
     \frac{d^2\psi}{dz^2}+p(z)\frac{d\psi}{dz}+q(z)\psi(z)=0 \;,\quad
     p(z)=\frac{f'(z)}{f(z)} \;,\quad
     q(z)=\frac{x_+^2\Omega^2-V(z)}{f^2(z)} \;.
\end{equation}
Since $p(z)$ and $q(z)$ have poles of order one and two at $z=1$, respectively, this point is a regular singular point of Eq.~\eqref{ODEZ} according to Frobenius theory \cite{ince1956ordinary}. Accordingly, the local solutions admit Frobenius expansions of the form
\begin{equation}
\psi(z)=(z-1)^\rho\sum_{\kappa=0}^\infty a_\kappa(z-1)^\kappa \;.
\end{equation}
The leading behavior at $z=1$ is represented by the factor $(z-1)^\rho$, where $\rho$ is determined by the indicial equation
\begin{equation}\label{indicial}
        \rho(\rho-1)+\mathcal P_0\rho+\mathcal Q_0=0
\end{equation}
with
\begin{equation}
        \mathcal P_0=\lim_{z\to1}(z-1)p(z)=1 \;, \qquad
        \mathcal Q_0=\lim_{z\to1}(z-1)^2q(z)=\left(\frac{x_+\Omega}{f'(1)}\right)^2 \;.
\end{equation}
The roots of Eq.~\eqref{indicial} are $\rho_\pm=\pm ix_+\Omega/f'(1)$. The negative-sign branch corresponds to purely ingoing behavior at the event horizon, so the QNM boundary condition is
\begin{equation}\label{QNMBCz1}
    \psi\underset{z\to1^+}{\longrightarrow}(z-1)^{-ix_+\alpha\Omega} \; , \qquad
    \alpha=\frac{1}{f'(1)}=\frac{1}{3x_+^2+1} \;.
\end{equation}

\item
{\textit{Asymptotic behavior as $z\to+\infty$}}: Unlike in asymptotically flat or de Sitter geometries, the AdS boundary is not a wave zone. Hence, no outgoing-wave condition is imposed at infinity. Instead, the asymptotic behavior is algebraic and is determined by the indicial structure of the radial equation. From Eq.~\eqref{fz}, we have
\begin{equation}
f(z)=x_+^2z^2+1-\frac{1+x_+^2}{z} \; ,
\end{equation}
and therefore
\begin{equation}
p(z)=\frac{2}{z}+\mathcal O\!\left(\frac{1}{z^3}\right) \; ,\qquad
q(z)=-\frac{2}{z^2}+\mathcal O\!\left(\frac{1}{z^4}\right) \; .
\end{equation}
Thus, the point at infinity is a regular singular point \cite{Bender1999}. Looking for solutions of the form $\psi(z)\sim z^\rho$ as $z\to+\infty$, one obtains the indicial equation $\rho^2+\rho-2=0$, whose roots are $\rho_1=-2$ and $\rho_2=1$. The two asymptotic branches are therefore
\begin{equation}\label{generalbehavior}
\psi(z)\underset{z\to+\infty}{\longrightarrow}Az+\frac{B}{z^2} \; .
\end{equation}
The standard vanishing-field condition at the AdS boundary sets $A=0$ \cite{HorowitzPRD2000,CardosoPRD2001,DaghighGreenMorey2023}, so that the QNM boundary condition at the conformal AdS boundary is
\begin{equation}\label{QNMBCs0inf}
\psi(z)\underset{z\to+\infty}{\longrightarrow}\frac{B}{z^2} \; .
\end{equation}
This asymptotic behavior defines the standard Dirichlet benchmark considered in the SAdS scalar QNM literature. The generalized coefficient boundary condition introduced below instead retains both asymptotic branches and imposes a relation between their coefficients.
\end{enumerate}
Having identified the asymptotic behavior at both boundaries, we introduce a new radial function $\Phi(z)$ by factoring out the ingoing horizon behavior and the Dirichlet falloff, so that $\Phi(z)$ is regular at both $z=1$ and the conformal AdS boundary
\begin{equation}\label{Ansatz}
\psi(z)=z^{-2+i\beta\Omega}(z-1)^{-i\beta\Omega}\Phi(z) \; ,\qquad
\beta=\frac{x_+}{3x_+^2+1}\;.
\end{equation}
Substituting Eq.~\eqref{Ansatz} into Eq.~\eqref{ourODE} yields the following ordinary differential equation for the regular radial function
\begin{equation}\label{ODEznone}
    P_2(z)\Phi''(z)+P_1(z)\Phi'(z)+P_0(z)\Phi(z)=0 \;,
\end{equation}
with
\begin{align}
    P_2(z)&=z^2(z-1)^2f^2(z) \;,\\
    P_1(z)&=(z-z^2)f(z)\left[(z-z^2)f'(z)+2f(z)(i\beta\Omega+2z-2)\right]\;,\\
    P_0(z)&=-\Omega^2Q_2(z)+i\Omega Q_1(z)+Q_0(z) \;,\label{P0z}\\
    Q_2(z)&=\beta^2f^2(z)-x_+^2z^2(z-1)^2,\qquad
    Q_1(z)=\beta(z-z^2)f(z)f'(z)+\beta f^2(z)(6z-5) \;,\\
    Q_0(z)&=-2z(z-1)^2f(z)f'(z)+6f^2(z)(z-1)^2-(z-z^2)^2V(z) \; .
\end{align}
We now compactify the radial domain by introducing the transformation
\begin{equation}\label{compactification}
z(y)=\frac{2}{1-y}\;,
\end{equation}
which maps the event horizon to $y=-1$ and the conformal AdS boundary to $y=1$. In the following, a dot denotes differentiation with respect to $y$. Under this transformation, Eq.~\eqref{ODEznone} becomes
\begin{equation}\label{ODEynone}
S_2(y)\ddot{\Phi}(y)+S_1(y)\dot{\Phi}(y)+S_0(y)\Phi(y)=0\;,
\end{equation}
where
\begin{align}
S_2(y)&=f^2(y)(1+y)^2\;,\label{S2}\\
S_1(y)&=-2i\beta\Omega(1+y)f^2(y)
-\frac{6(1+y)^2}{1-y}f^2(y)
+(1+y)^2f(y)\dot{f}(y)\;,\\
S_0(y)&=\Omega^2\Sigma_2(y)+i\Omega\Sigma_1(y)+\Sigma_0(y) \; ,
\end{align}
with
\begin{align}
\Sigma_2(y)&=\frac{4x_+^2(1+y)^2}{(1-y)^4}-\beta^2f^2(y) \; ,\nonumber\\
\Sigma_1(y)&=-\beta(1+y)f(y)\dot{f}(y)
+\frac{\beta(7+5y)}{1-y}f^2(y)\;,\\
\Sigma_0(y)&=-\frac{2(1+y)^2}{1-y}f(y)\dot{f}(y)
+2\left(\frac{1+y}{1-y}\right)^2
\left[3f^2(y)-\frac{2V(y)}{(1-y)^2}\right]\;.\label{Sigma0}
\end{align}
The regularized radial function $\Phi(y)$ must remain regular at $y=\pm1$. The transformed metric function and effective potential may be written as
\begin{eqnarray}
f(y)&=&(1+y)g(y)\; ,\qquad
g(y)=\frac{(x_+^2+1)y^2-2y(2x_+^2+1)+7x_+^2+1}
{2(1-y)^2}\; ,\label{fy}\\
V(y)&=&\frac{1}{4}(1-y)^2f(y)
\left[(1-y)\dot{f}(y)+\ell(\ell+1)\right]\;.\label{Vy}
\end{eqnarray}
\begin{table}[t]
\caption{\textit{Classification of the endpoint behavior of the functions defined in Eqs.~\eqref{S2}-\eqref{Vy}.}}
\label{tableEinsnone}
\centering
\small
\setlength{\tabcolsep}{6pt}
\renewcommand{\arraystretch}{1.25}
\begin{tabular}{cccccc}
\toprule
$y$ & $f(y)$ & $V(y)$ & $S_2(y)$ & $S_1(y)$ & $S_0(y)$\\
\hline
$-1$ & zero of order $1$ & zero of order $1$ & zero of order $4$ & zero of order $3$ & zero of order $3$\\
$+1$ & pole of order $2$ & pole of order $2$ & pole of order $4$ & pole of order $5$ & pole of order $5$\\
\hline
\end{tabular}
\end{table}
Table~\ref{tableEinsnone} shows that the coefficients of Eq.~\eqref{ODEynone} share a common zero of order $3$ at $y=-1$, while at $y=1$, $S_2(y)$ has a pole of order $4$ and $S_1(y)$ and $S_0(y)$ have poles of order $5$. To obtain coefficient functions that are regular at both endpoints, we multiply Eq.~\eqref{ODEynone} by the factor $(1-y)^5/(1+y)^3$. The resulting differential equation is
\begin{equation}\label{ODEhynone}
M_2(y)\ddot{\Phi}(y)+M_1(y)\dot{\Phi}(y)+M_0(y)\Phi(y)=0\;,
\end{equation}
where
\begin{eqnarray}
M_2(y)&=&(1-y)^5(1+y)g^2(y)\;,\label{M2}\\
M_1(y)&=&i\Omega N_1(y)+N_0(y)\;,\qquad
M_0(y)=\Omega^2C_2(y)+i\Omega C_1(y)+C_0(y)\;,\label{M10}
\end{eqnarray}
with
\begin{align}
N_1(y)&=-2\beta(1-y)^5g^2(y)\;,\label{N1}\\
N_0(y)&=-6(1-y)^4(1+y)g^2(y)
+(1-y)^5g(y)\left[(1+y)\dot{g}(y)+g(y)\right]\;,\label{N0}\\
C_2(y)&=\frac{1-y}{1+y}
\left[4x_+^2-\beta^2g^2(y)(1-y)^4\right]\;,\label{C2}\\
C_1(y)&=\beta\frac{(1-y)^4}{1+y}
\left\{(7+5y)g^2(y)
-(1-y)g(y)\left[(1+y)\dot{g}(y)+g(y)\right]\right\}\;,\label{C1}\\
C_0(y)&=-(1-y)^3g(y)
\left[3(1-y^2)\dot{g}(y)-3(3y+1)g(y)+\ell(\ell+1)\right]\;.\label{C0}
\end{align}
Evaluating the coefficient functions at the endpoints gives
\begin{align}
\lim_{y\to1^-}M_2(y)&=0=\lim_{y\to-1^+}M_2(y)\;,\label{lim2}\\
\lim_{y\to1^-}M_1(y)&=-32x_+^4 \; ,\qquad
\lim_{y\to-1^+}M_1(y)=iD_1\Omega+D_0\;,\\
\lim_{y\to1^-}M_0(y)&=\frac{16i\Omega x_+^5}{3x_+^2+1} \; ,\qquad
\lim_{y\to-1^+}M_0(y)=B_2\Omega^2+iB_1\Omega+B_0\;,\label{lim0}
\end{align}
where
\begin{align}
D_1&=-16x_+(3x_+^2+1) \; ,\qquad
D_0=8(3x_+^2+1)^2\;,\label{D01}\\
B_2&=\frac{8x_+^2(3x_+^2+2)}{3x_+^2+1} \; ,\qquad
B_1=12x_+(5x_+^2+2) \; ,\nonumber\\
B_0&=-4(3x_+^2+1)
\left[\ell(\ell+1)+3(3x_+^2+1)\right]\;.\label{B012}
\end{align}
Finally, we rewrite Eq.~\eqref{ODEhynone} in the quadratic operator-pencil form
\begin{equation}\label{TSCH}
L_0\left[\Phi,\dot{\Phi},\ddot{\Phi}\right]
+iL_1\left[\Phi,\dot{\Phi},\ddot{\Phi}\right]\Omega
+L_2\left[\Phi,\dot{\Phi},\ddot{\Phi}\right]\Omega^2=0 \; ,
\end{equation}
with
\begin{align}
L_0\left[\Phi,\dot{\Phi},\ddot{\Phi}\right]
&=L_{00}(y)\Phi+L_{01}(y)\dot{\Phi}+L_{02}(y)\ddot{\Phi}\;,\label{L0none}\\
L_1\left[\Phi,\dot{\Phi},\ddot{\Phi}\right]
&=L_{10}(y)\Phi+L_{11}(y)\dot{\Phi}+L_{12}(y)\ddot{\Phi}\;,\label{L1none}\\
L_2\left[\Phi,\dot{\Phi},\ddot{\Phi}\right]
&=L_{20}(y)\Phi+L_{21}(y)\dot{\Phi}+L_{22}(y)\ddot{\Phi}\;.\label{L2none}
\end{align}
Table~\ref{tableZweinone} summarizes the coefficients $L_{ij}$ appearing in Eqs.~\eqref{L0none}--\eqref{L2none}, together with their limiting values at $y=\pm1$.
\begin{table}[t]
\caption{\textit{Definitions of the coefficients $L_{ij}$ appearing in Eqs.~\eqref{L0none}-\eqref{L2none} together with their limiting values at the endpoints of the computational interval. The functions $N_i$, $C_i$, and $M_2$ are defined in Eqs.~\eqref{N1}-\eqref{C0} and \eqref{M2}.}}
\label{tableZweinone}
\centering
\small
\renewcommand{\arraystretch}{1.25}
\setlength{\tabcolsep}{8pt}
\begin{tabular}{cccc}
\toprule
$(i,j)$ &
$\displaystyle\lim_{y\rightarrow-1^+}L_{ij}$ &
$L_{ij}$ &
$\displaystyle\lim_{y\rightarrow1^-}L_{ij}$ \\
\midrule
$(0,0)$ & $B_0$ & $C_0$ & $0$ \\
$(0,1)$ & $D_0$ & $N_0$ & $-32x_+^4$ \\
$(0,2)$ & $0$ & $M_2$ & $0$ \\
\midrule
$(1,0)$ & $B_1$ & $C_1$ & $\displaystyle\frac{16x_+^5}{3x_+^2+1}$ \\
$(1,1)$ & $D_1$ & $N_1$ & $0$ \\
$(1,2)$ & $0$ & $0$ & $0$ \\
\midrule
$(2,0)$ & $B_2$ & $C_2$ & $0$ \\
$(2,1)$ & $0$ & $0$ & $0$ \\
$(2,2)$ & $0$ & $0$ & $0$ \\
\bottomrule
\end{tabular}
\end{table}

\subsubsection{The scalar case with generalized coefficient boundary conditions at the conformal AdS boundary}

The previous subsection considered the standard vanishing-field condition at the conformal AdS boundary, corresponding to the removal of the slow asymptotic branch of the minimally coupled massless scalar field. In the present subsection, we consider instead a one-parameter family of generalized coefficient boundary conditions relating the two asymptotic coefficients of the same massless scalar field. We stress that, for the minimally coupled massless scalar, this should be viewed as a deformation of the spectral problem rather than as the usual alternative quantization available for scalars in the Breitenlohner-Freedman window \cite{IshibashiWald2004,Kinoshita2024}. In particular, the standard finite-energy prescription selects the Dirichlet branch, whereas the mixed conditions introduced below are imposed as generalized boundary conditions on the QNM eigenvalue problem. These conditions should therefore be regarded as a deformation of the massless scalar spectral problem, not as the standard Robin family associated with alternative quantization. As before, the boundary condition at the event horizon is purely ingoing. The essential difference from the Dirichlet formulation is that the slow asymptotic branch is retained throughout the calculation. Rather than removing it {\it a priori}, the relation between the slow and fast asymptotic coefficients is imposed directly as a boundary condition on the spectral eigenvalue problem. We again start from Eq.~\eqref{ourODE}, and determine the asymptotic behavior of the radial function $\psi$ in the two relevant regions. For $z\to1^+$, the analysis is unchanged and the ingoing QNM condition is therefore given by Eq.~\eqref{QNMBCz1}. As $z\to+\infty$, we have the two independent asymptotic branches given by Eq.~\eqref{generalbehavior}. The standard Dirichlet condition sets $A=0$. In the present generalized problem, we instead impose a linear relation between $A$ and $B$. To express this relation in terms of the scalar field itself, we use Eq.~\eqref{generalbehavior} together with the rescalings $z=r/r_+$ and $x_+=r_+/R$. Then, a straightforward computation gives
\begin{equation}
\frac{\psi(z)}{z}\underset{z\to+\infty}{\longrightarrow}
A+\frac{B}{z^3}\; .
\end{equation}
Thus, the two independent coefficients of the scalar field are naturally identified with $A$ and $B$. We impose the one-parameter family of generalized coefficient boundary conditions
\begin{equation}\label{thetaBCphi}
A\cos\theta+B\sin\theta=0 \; ,\qquad
0\leq\theta\leq\frac{\pi}{2} \; .
\end{equation}
The endpoint $\theta=0$ gives the standard Dirichlet condition $A=0$, while $\theta=\pi/2$ gives the coefficient-Neumann condition $B=0$. We shall refer to intermediate values of $\theta$ as generalized coefficient boundary conditions. These conditions should not be confused with the standard Robin family arising from alternative quantization for scalar fields in the Breitenlohner-Freedman window. Note that it is important to emphasize that Eq.~\eqref{thetaBCphi} does not arise from introducing a different scalar theory. Rather, it provides a one-parameter deformation of the spectral problem for the same minimally coupled massless scalar field considered throughout this work. In order to construct an ansatz that incorporates the boundary condition \eqref{thetaBCphi} together with the correct asymptotic behavior at the event horizon and at the AdS boundary, we consider the following factorization ansatz
\begin{equation}\label{mixedAnsatz}
\psi(z)=z\,\mathcal H(z;\Omega)\,\chi(z) \; ,
\end{equation}
where
\begin{equation}\label{horizonFactorMixed}
\mathcal H(z;\Omega)=e^{-ix_+\alpha\Omega\left[\ln{\left(1-\frac{1}{z}\right)}+\frac{1}{z}+\frac{1}{2z^2}+\frac{1}{3z^3}\right]} \; .
\end{equation}
We further require that $\chi(z)$ remains regular at the event horizon and throughout the exterior region. The factor $\mathcal H(z;\Omega)$ has two useful properties. Near the event horizon, $z\to1^+$, it behaves as
\begin{equation}
\mathcal H(z;\Omega)\underset{z\to1^+}{\longrightarrow}(z-1)^{-ix_+\alpha\Omega} \; ,
\end{equation}
and hence Eq.~\eqref{mixedAnsatz} automatically implements the ingoing QNM behavior. Near the conformal AdS boundary,
$z\to+\infty$, one has
\begin{equation}
\mathcal H(z;\Omega)=1+\mathcal O\!\left(\frac{1}{z^4}\right) \; .
\end{equation}
We therefore write the regular function $\chi(z)$ as
\begin{equation}\label{expchi}
\chi(z)=A+\frac{\mathcal C_2}{z^2}+\frac{B}{z^3}
+\mathcal O\!\left(\frac1{z^4}\right) \; ,
\qquad z\to+\infty \; .
\end{equation}
The coefficient of $z^{-1}$ vanishes, while $\mathcal C_2$ is fixed by the differential equation. For the massless scalar one finds
\begin{equation}
\mathcal C_2=\frac{\Omega^2-\ell(\ell+1)}{2x_+^2}\,A \; .
\end{equation}
This implies that
\begin{equation}
\frac{\psi(z)}{z}
=
A+\frac{\mathcal C_2}{z^2}
+\frac{B}{z^3}
+\mathcal O\!\left(\frac1{z^4}\right) \; ,
\end{equation}
and the coefficients $A$ and $B$ appearing in Eq.~\eqref{thetaBCphi} are therefore encoded directly in the asymptotic expansion of $\chi(z)$. Substitution of Eq.~\eqref{mixedAnsatz} into Eq.~\eqref{ourODE} gives
\begin{equation}\label{ODEzmixed}
\widetilde P_2(z)\chi''(z)
+\widetilde P_1(z)\chi'(z)
+\widetilde P_0(z)\chi(z)=0 \; ,
\end{equation}
where primes denote differentiation with respect to $z$, and
\begin{align}
\widetilde P_2(z)&=z^8(z-1)^2f^2(z)\;,\\
\widetilde P_1(z)&=z^4(z-1)f(z)
\left[z^4(z-1)f'(z)
+2f(z)(z^4-z^3-i\beta\Omega)\right]\;,\\
\widetilde P_0(z)&=-\Omega^2\widetilde Q_2(z)
+i\Omega\widetilde Q_1(z)
+\widetilde Q_0(z)\;,\label{P0z2}\\
\widetilde Q_2(z)&=\beta^2f^2(z)-x_+^2z^8(z-1)^2,\quad
\widetilde Q_1(z)=-\beta z^4(z-1)f(z)f'(z)+\beta z^3f^2(z)(3z-2)\;,\\
\widetilde Q_0(z)&=z^7(z-1)^2f(z)f'(z)-z^8(z-1)^2V(z) \; .
\end{align}
We compactify the exterior region using the same transformation \eqref{compactification}. The event horizon is mapped to $y=-1$, while the conformal AdS boundary is mapped to $y=1$. As before, a dot denotes differentiation with respect to $y$. Eq.~\eqref{ODEzmixed} then becomes
\begin{equation}\label{ODEymixed}
\widetilde S_2(y)\ddot\chi(y)
+\widetilde S_1(y)\dot\chi(y)
+\widetilde S_0(y)\chi(y)=0 \; ,
\end{equation}
where
\begin{align}
\widetilde S_2(y)&=\frac{64f^2(y)(1+y)^2}{(1-y)^6}\;,\label{S2bc}\\
\widetilde S_1(y)&=-\frac{128f^2(y)(1+y)^2}{(1-y)^7}
+\frac{8(1+y)f(y)}{(1-y)^7}
\left\{
8(1-y^2)\dot f(y)
+2f(y)\left[16-8(1-y)-i\beta\Omega(1-y)^4\right]
\right\}\;,\\
\widetilde S_0(y)&=\Omega^2\widetilde\Sigma_2(y)
+i\Omega\widetilde\Sigma_1(y)
+\widetilde\Sigma_0(y) \; ,
\end{align}
with
\begin{align}
\widetilde\Sigma_2(y)&=
\frac{256(1+y)^2x_+^2}{(1-y)^{10}}
-\beta^2f^2(y) \; ,\quad
\widetilde\Sigma_1(y)=
-\frac{8\beta(1+y)}{(1-y)^3}f(y)\dot f(y)
+\frac{16\beta(2+y)}{(1-y)^4}f^2(y)\;,\\
\widetilde\Sigma_0(y)&=
\frac{64(1+y)^2}{(1-y)^7}f(y)\dot f(y)
-\frac{256(1+y)^2}{(1-y)^{10}}V(y)\;.\label{Sigma0bc}
\end{align}
In practice, Eq.~\eqref{ODEymixed} is imposed only at the interior collocation points. Furthermore, the functions $f(y)$ and $V(y)$ are given by Eqs.~\eqref{fy} and \eqref{Vy}, respectively.
\begin{table}[t]
\caption{\textit{Classification of the endpoint behavior of the coefficient functions defined in Eqs.~\eqref{S2bc}-\eqref{Sigma0bc}.}}
\label{tableEinsnonebc}
\centering
\small
\setlength{\tabcolsep}{6pt}
\renewcommand{\arraystretch}{1.25}
\begin{tabular}{cccccc}
\toprule
$y$ & $f(y)$ & $V(y)$ & $\widetilde S_2(y)$ & $\widetilde S_1(y)$ & $\widetilde S_0(y)$\\
\midrule
$-1$ & zero of order $1$ & zero of order $1$ & zero of order $4$ & zero of order $3$ & zero of order $3$\\
$+1$ & pole of order $2$ & pole of order $2$ & pole of order $10$ & pole of order $11$ & pole of order $10$\\
\bottomrule
\end{tabular}
\end{table}
Table~\ref{tableEinsnonebc} shows that the coefficients of Eq.~\eqref{ODEymixed} share a common zero of order $3$ at $y=-1$, while at $y=1$, $\widetilde S_1(y)$ has a pole of order $11$ and $\widetilde S_2(y)$ and $\widetilde S_0(y)$ have poles of order $10$. To obtain coefficient functions that are regular at both endpoints, we multiply Eq.~\eqref{ODEymixed} by the factor $(1-y)^{11}/(1+y)^3$. The resulting differential equation is
\begin{equation}\label{ODEhynoneM}
\widetilde M_2(y)\ddot{\chi}(y)
+\widetilde M_1(y)\dot{\chi}(y)
+\widetilde M_0(y)\chi(y)=0\;,
\end{equation}
where
\begin{align}
\widetilde M_2(y)&=64(1-y)^5(1+y)g^2(y)\;,\label{Mt2}\\
\widetilde M_1(y)&=i\Omega\widetilde N_1(y)+\widetilde N_0(y)\;,\qquad
\widetilde M_0(y)=\Omega^2\widetilde C_2(y)+i\Omega\widetilde C_1(y)+\widetilde C_0(y)\;.\label{M10t}
\end{align}
Here
\begin{align}
\widetilde N_1(y)&=-16\beta(1-y)^8g^2(y)\;,\qquad
\widetilde N_0(y)=64(1-y)^5g(y)\left[g(y)+(1+y)\dot{g}(y)\right]\;,\label{Nt01}\\
\widetilde C_2(y)&=\frac{1-y}{1+y}
\left[256x_+^2-\beta^2g^2(y)(1-y)^{10}\right]\;,\label{Ct2}\\
\widetilde C_1(y)&=8\beta g(y)(1-y)^7
\left[3g(y)-(1-y)\dot{g}(y)\right]\;,\qquad
\widetilde C_0(y)=-64\ell(\ell+1)g(y)(1-y)^3\;.\label{Ct01}
\end{align}
Evaluating the coefficient functions at the endpoints gives
\begin{align}
\lim_{y\to1^-}\widetilde M_2(y)
&=0=\lim_{y\to-1^+}\widetilde M_2(y)\;,\label{limt2}\\
\lim_{y\to1^-}\widetilde M_1(y)
&=1024x_+^4 \; ,\qquad
\lim_{y\to-1^+}\widetilde M_1(y)
=i\widetilde D_1\Omega+\widetilde D_0\;,\\
\lim_{y\to1^-}\widetilde M_0(y)
&=0 \; ,\qquad
\lim_{y\to-1^+}\widetilde M_0(y)
=\widetilde B_2\Omega^2+i\widetilde B_1\Omega+\widetilde B_0\;.\label{limt0}
\end{align}
The constants appearing in these limits are
\begin{align}
\widetilde D_1&=-1024x_+(3x_+^2+1)\; ,\qquad
\widetilde D_0=512(3x_+^2+1)^2\;,\label{Dt01}\\
\widetilde B_2&=\frac{512x_+^2(12x_+^2+5)}{3x_+^2+1}\; ,\qquad
\widetilde B_1=768x_+(2x_+^2+1)\; ,\nonumber\\
\widetilde B_0&=-256\ell(\ell+1)(3x_+^2+1)\;.\label{Bt012}
\end{align}
Finally, we rewrite Eq.~\eqref{ODEhynoneM} in the quadratic operator-pencil form
\begin{equation}\label{TSCH2}
\widetilde L_0\left[\chi,\dot{\chi},\ddot{\chi}\right]
+i\widetilde L_1\left[\chi,\dot{\chi},\ddot{\chi}\right]\Omega
+\widetilde L_2\left[\chi,\dot{\chi},\ddot{\chi}\right]\Omega^2=0 \; ,
\end{equation}
with
\begin{align}
\widetilde L_0\left[\chi,\dot{\chi},\ddot{\chi}\right]
&=\widetilde L_{00}(y)\chi
+\widetilde L_{01}(y)\dot{\chi}
+\widetilde L_{02}(y)\ddot{\chi}\;,\label{L0tnone}\\
\widetilde L_1\left[\chi,\dot{\chi},\ddot{\chi}\right]
&=\widetilde L_{10}(y)\chi
+\widetilde L_{11}(y)\dot{\chi}
+\widetilde L_{12}(y)\ddot{\chi}\;,\label{L1tnone}\\
\widetilde L_2\left[\chi,\dot{\chi},\ddot{\chi}\right]
&=\widetilde L_{20}(y)\chi
+\widetilde L_{21}(y)\dot{\chi}
+\widetilde L_{22}(y)\ddot{\chi}\;.\label{L2tnone}
\end{align}
Table~\ref{tableZweinone2} summarizes the coefficients $\widetilde L_{ij}$ appearing in Eqs.~\eqref{L0tnone}-\eqref{L2tnone}, together with their limiting values at $y=\pm1$.
\begin{table}[t]
\caption{\textit{Definitions of the coefficients $\widetilde L_{ij}$ appearing in Eqs.~\eqref{L0tnone}-\eqref{L2tnone}, together with their limiting values at the endpoints of the computational interval. The functions $\widetilde N_i$, $\widetilde C_i$, and $\widetilde M_2$ are defined in Eqs.~\eqref{Mt2}-\eqref{Ct01}.}}
\label{tableZweinone2}
\centering
\small
\renewcommand{\arraystretch}{1.25}
\setlength{\tabcolsep}{8pt}
\begin{tabular}{cccc}
\toprule
$(i,j)$ &
$\displaystyle\lim_{y\to-1^+}\widetilde L_{ij}$ &
$\widetilde L_{ij}$ &
$\displaystyle\lim_{y\to1^-}\widetilde L_{ij}$\\
\midrule
$(0,0)$ & $\widetilde B_0$ & $\widetilde C_0$ & $0$\\
$(0,1)$ & $\widetilde D_0$ & $\widetilde N_0$ & $1024x_+^4$\\
$(0,2)$ & $0$ & $\widetilde M_2$ & $0$\\
\midrule
$(1,0)$ & $\widetilde B_1$ & $\widetilde C_1$ & $0$\\
$(1,1)$ & $\widetilde D_1$ & $\widetilde N_1$ & $0$\\
$(1,2)$ & $0$ & $0$ & $0$\\
\midrule
$(2,0)$ & $\widetilde B_2$ & $\widetilde C_2$ & $0$\\
$(2,1)$ & $0$ & $0$ & $0$\\
$(2,2)$ & $0$ & $0$ & $0$\\
\bottomrule
\end{tabular}
\end{table}
Furthermore, the endpoint $y=1$ is used to impose the generalized coefficient boundary condition \eqref{thetaBCphi}. To express Eq.~\eqref{thetaBCphi} in terms of $\chi(y)$, we use Eq.~\eqref{compactification} together with Eq.~\eqref{expchi}. The resulting relations between the asymptotic coefficients and the endpoint data are
\begin{equation}
A=\chi(1) \; ,\qquad
B=-\frac{4}{3}\chi^{(3)}(1) \; ,
\end{equation}
where the third derivative is taken with respect to $y$. The generalized coefficient boundary condition \eqref{thetaBCphi} therefore becomes
\begin{equation}\label{thetaBCchi}
\cos\theta\,\chi(1)
-\frac{4}{3}\sin\theta\,\chi^{(3)}(1)=0 \; .
\end{equation}
Thus, $\theta=0$ implies $\chi(1)=0$, which reproduces the standard Dirichlet condition, while $\theta=\pi/2$ requires $\chi^{(3)}(1)=0$, corresponding to the coefficient-Neumann condition. Finally, we expand the regular function $\chi(y)$ in Chebyshev polynomials,
\begin{equation}\label{ChebMixed}
\chi(y)=\sum_{j=0}^{N-1}b_jT_j(y) \; .
\end{equation}
Using the identities \cite{Abramowitz1972}
\begin{equation}
T_j(1)=1 \; ,\qquad
T_j^{(3)}(1)=\frac{j^2(j^2-1)(j^2-4)}{15} \; ,
\end{equation}
the boundary condition \eqref{thetaBCchi} yields the spectral boundary row
\begin{equation}\label{spectralBCrowMixed}
\sum_{j=0}^{N-1}
\left[
\cos\theta
-\frac{4j^2(j^2-1)(j^2-4)}{45}\sin\theta
\right]b_j=0.
\end{equation}
The remaining $N-1$ rows are obtained by imposing the regularized differential equation \eqref{ODEhynoneM} at the interior collocation points. Since the differential-equation rows are quadratic in $\Omega$, while the boundary row \eqref{spectralBCrowMixed} is independent of $\Omega$, the final discretized problem remains quadratic in $\Omega$. The Dirichlet endpoint $\theta=0$ provides a direct consistency check because it must reproduce the standard scalar spectrum computed in the previous subsection. For $\theta\neq0$, the spectrum describes a generalized coefficient deformation of the massless scalar QNM problem.


\section{Spectral method and extraction of the QNM spectrum}\label{SecSpectralMethod}

In this section, we describe the numerical implementation used to compute the scalar QNM spectra discussed below. The method is based on a Chebyshev collocation discretization of the regularized radial equations derived in the previous section. In all cases, the final algebraic problem is a quadratic matrix pencil in the dimensionless frequency $\Omega=\omega R$, namely
\begin{equation}\label{QEPgeneral}
\left(\mathcal M_0+i\Omega \mathcal M_1+\Omega^2\mathcal M_2\right)\mathbf b=0 \; ,
\end{equation}
where $\mathbf b=(b_0,\ldots,b_{N-1})^T$ contains the Chebyshev expansion coefficients of the regular part of the radial eigenfunction. After compactifying the exterior region according to Eq.~\eqref{compactification}, the event horizon is located at $y=-1$, while the conformal AdS boundary is situated at $y=1$. The regular radial function is expanded as
\begin{equation} \label{ChebExpansionMethod}
    F_N(y)=\sum_{j=0}^{N-1} b_j T_j(y) \; ,
\end{equation}
where $T_j(y)$ denotes the Chebyshev polynomial of degree $j$. This is the standard Chebyshev collocation representation used in spectral methods for boundary-value and eigenvalue problems \cite{Boyd2000}. The radial equation is then imposed at Chebyshev-Gauss collocation points. For a problem in which both boundary conditions are already built into the ansatz, we use $N$ collocation points,
\begin{equation}\label{ChebGaussNodes}
y_i=\cos\left(\frac{(2i-1)\pi}{2N}\right),\qquad i=1,\ldots,N \; .
\end{equation}
At each point $y_i$, the equation has the schematic form
\begin{equation}\label{SpectralOperator}
L_0[F_N,\dot F_N,\ddot F_N](y_i)+i\Omega L_1[F_N,\dot F_N,\ddot F_N](y_i) +\Omega^2 L_2[F_N,\dot F_N,\ddot F_N](y_i)=0 \; ,
\end{equation}
where the dot denotes differentiation with respect to $y$. Since $F_N,\dot F_N,\ddot F_N$ are linear in the coefficients $b_j$, Eq.~\eqref{SpectralOperator} generates one row of the matrices $\mathcal M_0$, $\mathcal M_1$, $\mathcal M_2$. Imposing the equation at all collocation points yields the quadratic pencil \eqref{QEPgeneral}. For the standard massless scalar problem, the QNM boundary conditions are purely ingoing behavior at the event horizon and the vanishing-field condition at the conformal AdS boundary. As shown above, these conditions are implemented by means of the ansatz \eqref{Ansatz}. Notice that the prefactor in Eq.~\eqref{Ansatz} contains the ingoing horizon behavior and the standard scalar falloff $\psi\sim z^{-2}$ at the AdS boundary. Therefore, the remaining function $\Phi$ is regular at both endpoints. After the transformation \eqref{compactification}, we expand
\begin{equation}
\Phi_N(y)=\sum_{j=0}^{N-1}b_jT_j(y) \; ,
\end{equation}
and impose the regularized equation at the $N$ points \eqref{ChebGaussNodes}. This gives an $N\times N$ quadratic pencil of the form \eqref{QEPgeneral}. For the generalized coefficient boundary condition, the AdS-boundary condition is not built into the ansatz. Instead, we factor out only the ingoing behavior at the horizon, while keeping both independent boundary coefficients available. The ansatz is provided by \eqref{mixedAnsatz} together with \eqref{horizonFactorMixed}, while the regular function $\chi(z)$ has the expansion \eqref{expchi}, where the coefficients $A$ and $B$ are the two independent boundary data, which satisfy the constraint \eqref{thetaBCphi}. The endpoint $\theta=0$ gives the usual Dirichlet condition $A=0$, while $\theta=\pi/2$ gives the coefficient-Neumann condition $B=0$. In the previous section, we showed that this constraint can be expressed in terms of $\chi(y)$ as in \textcolor{blue}{Eq.} \eqref{thetaBCchi}. For the Chebyshev expansion \eqref{ChebMixed}, this boundary condition is represented by the spectral row \eqref{spectralBCrowMixed}. The generalized-boundary calculation is assembled as follows. We impose the regularized differential equation at $N-1$ Chebyshev--Gauss points
\begin{equation}\label{InteriorNodesMixedMethod}
y_i=\cos\left(\frac{(2i-1)\pi}{2(N-1)}\right),\qquad i=1,\ldots,N-1 \;,
\end{equation}
and use Eq.~\eqref{spectralBCrowMixed} as the final row. Hence, the matrices again have size $N\times N$, but their last row is a boundary-condition row rather than a collocation row. Since Eq.~\eqref{spectralBCrowMixed} is independent of $\Omega$, the last row contributes only to $\mathcal M_0$. More precisely, we have
\begin{equation}\label{BoundaryMatrixRow}
(\mathcal M_0)_{N,j+1}=\cos\theta-\frac{4}{45}\sin\theta\,j^2(j^2-1)(j^2-4) \; , \qquad
(\mathcal M_1)_{N,j+1}=(\mathcal M_2)_{N,j+1}=0 \; ,
\end{equation}
with $j\in\{0,\ldots,N-1\}$. In this formulation, the leading matrix $\mathcal M_2$ is singular because the boundary row is independent of $\Omega$. For the numerical solution, we therefore eliminate the boundary constraint before linearizing the quadratic pencil. To this end, let $c^T\mathbf b=0$ denote the discrete boundary condition \eqref{spectralBCrowMixed}, and let $Z$ be a basis of the nullspace of $c^T$. We write $\mathbf b=Z\mathbf u$. The first $N-1$ rows are the differential equation rows. If $\mathcal R$ denotes the matrix that selects these rows, the reduced quadratic pencil is
\begin{equation}\label{ReducedMixedPencil}
\left(\widehat{\mathcal{M}_0}+i\Omega\widehat{\mathcal{M}_1}+\Omega^2\widehat{\mathcal{M}_2}\right)\mathbf u=0 \; ,\qquad
\widehat{\mathcal{M}_k}=\mathcal{R} \mathcal{M}_k Z \; ,\qquad k\in\{0,1,2\} \; .
\end{equation}
This reduced pencil has dimension $(N-1)\times(N-1)$ and contains the boundary condition exactly at the discrete level. We emphasize that the null-space reduction is a feature of the generalized coefficient problem only. It is required because the spectral boundary row is independent of $\Omega$, so the corresponding row of the leading quadratic matrix vanishes.  Such a boundary-row formulation is one of the advantages of the spectral approach. Once the radial equation has been regularized, changing the boundary condition at the conformal AdS boundary amounts to changing the final spectral row, while the bulk collocation discretization is left unchanged. Thus, Dirichlet, coefficient-Neumann, Robin-type, or more general linear coefficient
conditions can be implemented within the same matrix-pencil framework. Note that for the standard Dirichlet problem no such reduction is necessary: the horizon and AdS-boundary behaviors are built into the ansatz, and the resulting $N\times N$ quadratic pencil is linearized directly. Since both boundary-value problems reduce to a quadratic pencil of the form \eqref{QEPgeneral}, we employ the companion linearization
\begin{equation}\label{CompanionLinearization}
\begin{pmatrix}
0 & I\\
-\mathcal M_0 & -i\mathcal M_1
\end{pmatrix}
\begin{pmatrix}
\mathbf b\\
\Omega\mathbf b
\end{pmatrix}
=
\Omega
\begin{pmatrix}
I & 0\\
0 & \mathcal M_2
\end{pmatrix}
\begin{pmatrix}
\mathbf b\\
\Omega\mathbf b
\end{pmatrix} \; .
\end{equation}
For the generalized-boundary problem, the same linearization is applied to the reduced matrices $\widehat{\mathcal{M}_k}$ in Eq.~\eqref{ReducedMixedPencil}. All matrix entries are generated symbolically in Maple, exported with high-precision numerical coefficients, and the resulting generalized eigenvalue problem is solved in Julia using arbitrary-precision arithmetic. For the calculations reported in the next section, we use $200$ decimal digits for the matrix entries and a working precision of $220$ decimal digits. Since the time dependence is $e^{-i\omega t}$, $\operatorname{Im}\Omega<0$ corresponds to damped modes, while $\operatorname{Im}\Omega>0$ corresponds to unstable modes. We do not discard modes with positive imaginary part. Instead, they are kept in the candidate set and classified separately. The classification tolerance is $\tau_{\rm damp}=10^{-18}$. Modes with $|\Omega|<10^{-30}$ are treated as static zero modes and are not counted as QNMs. Purely imaginary modes are identified by the criterion $|\operatorname{Re}\Omega|<10^{-10}$.

Moreover, for each choice of $x_+$, $\ell$, and boundary parameter $\theta$, we compute the spectrum at three spectral resolutions. In the production runs reported in this work, we use $N_1=190$, $N_2=195$, and $N_3=200$. Let $\Omega^{(1)}$, $\Omega^{(2)}$, and $\Omega^{(3)}$ denote matched modes at the three resolutions. The matching is performed by a nearest-neighbour search in the complex $\Omega$-plane. Starting from a candidate mode at $N_1$, we find the closest unused candidate at $N_2$, and then the closest unused candidate at $N_3$. We define
\begin{equation}
d_{12}=\left|\Omega^{(1)}-\Omega^{(2)}\right| \; ,\qquad
d_{23}=\left|\Omega^{(2)}-\Omega^{(3)}\right| \; ,\qquad
d_{13}=\left|\Omega^{(1)}-\Omega^{(3)}\right| \; .
\end{equation}
The mode is accepted if
\begin{equation}\label{TripletCriterion}
\max(d_{12},d_{23},d_{13})\leq\max\left[\tau_{\rm abs},\tau_{\rm rel}\max\left(1,|\Omega^{(1)}|,|\Omega^{(2)}|,|\Omega^{(3)}|\right)\right] \; ,
\end{equation}
with $\tau_{\rm abs}=10^{-4}$ and $\tau_{\rm rel}=10^{-4}$. The quoted value of the frequency is the arithmetic average
\begin{equation}\label{AveragedMode}
\Omega_{\rm avg}=\frac{\Omega^{(1)}+\Omega^{(2)}+\Omega^{(3)}}{3} \;.
\end{equation}
For each accepted triplet we also report the resolution-spread
\begin{equation}\label{MaxError}
\Delta\Omega_{\rm max}=\max(d_{12},d_{23},d_{13}) \; .
\end{equation}
This quantity is an internal convergence diagnostic: it measures the largest pairwise difference between the three matched frequencies obtained at $N\in\{190,195,200\}$. It is not a rigorous bound on the exact numerical error, but it provides a practical measure of spectral stability. For the accepted modes, the rapid decrease of the resolution-spread with increasing Chebyshev resolution is consistent with the spectral convergence
expected for the regularized eigenvalue problem. We nevertheless use the triplet-matching criterion as a conservative practical filter, rather than claiming a rigorous exponential-convergence estimate for each individual mode. Only modes that pass the triplet criterion \eqref{TripletCriterion} are used in the final tables and plots. When displaying symmetric oscillatory pairs, we usually list only the branch with positive real part, since the radial equation has real coefficients and the companion branch is obtained under the transformation $\Omega\mapsto -\Omega^*$. Throughout the numerical analysis we write $\Omega=\Omega_R+i\Omega_I$, with $\Omega_R=\operatorname{Re}\Omega$ and
$\Omega_I=\operatorname{Im}\Omega$. We use $\Re\Omega$ and $\Im\Omega$
interchangeably with $\operatorname{Re}\Omega$ and $\operatorname{Im}\Omega$
in figure labels and table headings.


\section{Numerical results for the SAdS\label{sec:results}}

We now present the numerical spectra obtained with the Chebyshev spectral method described above. The discussion is divided into two parts. First, we study the standard Dirichlet boundary condition at the conformal AdS boundary, which provides a direct benchmark against the existing Schwarzschild--AdS scalar QNM literature. Second, we investigate the generalized coefficient condition $A\cos\theta+B\sin\theta=0$, where the Dirichlet problem is recovered at $\theta=0$, and analyze the boundary-condition-induced instability that appears for nonzero $\theta$.

\subsection{The scalar sector with Dirichlet boundary condition at the conformal AdS boundary}

We begin with the standard massless scalar problem, imposing purely ingoing behavior at the event horizon and the Dirichlet, or vanishing-field, condition at the conformal AdS boundary. This case provides the natural benchmark for the spectral method, since it has been studied extensively with independent numerical techniques. Throughout this subsection we display frequencies in units of $\Omega=\omega R$, with the convention $\Omega=\Omega_R+i\Omega_I$, so that damped modes have $\Omega_I<0$. Unless otherwise stated, only the branch with positive real part is tabulated, the negative-real branch being obtained from the symmetry $\Omega\mapsto-\Omega^*$.

Table~\ref{tables0ell0n0} provides a first benchmark of our implementation for the fundamental scalar mode with $\ell=0$. The agreement with the Horowitz-Hubeny power-series results is excellent over the full range of black hole sizes considered. In particular, our values reproduce the quoted frequencies of Ref. \cite{HorowitzPRD2000} to all displayed digits, from the intermediate regime $x_+\sim1$ up to the large-black-hole regime $x_+\gg1$. This confirms that the compactification, the factorization of the horizon and AdS-boundary asymptotics, and the resulting quadratic matrix pencil reproduce the standard scalar SAdS QNM problem. The comparison with the interpolation matrix method of Ref. \cite{LinCQG2017} is also very good for $x_+\geq0.6$. The only visible discrepancy occurs at $x_+=0.4$, where the value reported in Ref. \cite{LinCQG2017} differs from both the Horowitz-Hubeny result and our spectral result. Since our value agrees with Ref. \cite{HorowitzPRD2000} at this point, this isolated
difference is consistent with either finite interpolation accuracy or a
discrepancy in the tabulated value of Ref.~\cite{LinCQG2017}, rather than with
a change in the boundary-value problem. Apart from this point, the three methods give mutually consistent results. The table also illustrates the expected large-black-hole scaling: for $x_+=50$ and $x_+=100$, the ratio $\Omega/x_+$ is already approximately constant, $\Omega/x_+\simeq1.85-2.66i$, showing the onset of the linear scaling of scalar SAdS QNMs with the horizon radius.

A more stringent comparison is given in Table~\ref{tables0ell0xp1}, where we list a long sequence of overtones for the intermediate-size black hole $x_+=1$ with $\ell=0$. The agreement with the data of Ref. \cite{CardosoKonoplyaLemos2003} and with the asymptotic iteration method of Ref. \cite{ChoCornellDoukasNaylor2010} is again excellent. For the modes $n\in\{0,1,2\}$ and $n\geq4$, our spectral values reproduce the quoted frequencies to all displayed digits and agree with the AIM values wherever those are available. The only apparent exception concerns the real part of the $n=3$ entry in the data quoted in Ref. \cite{CardosoKonoplyaLemos2003}. This mismatch was already noted in Ref. \cite{ChoCornellDoukasNaylor2010}, where the corrected value was obtained independently using the Mathematica notebook associated with the QNM data of Ref. \cite{Berti2009CQG}. The AIM value $\Omega_{n=3}=8.68233-9.74854i$ agrees with our spectral result, $\Omega_{n=3}=8.682227-9.748517i$. Thus, following the discussion in Ref.~\cite{ChoCornellDoukasNaylor2010}, we treat this entry as a previously identified mismatch in the quoted data rather than as a disagreement between the numerical methods.

Table~\ref{tables0ell0xp1} also shows that the overtones rapidly approach an almost uniformly spaced pattern. Using the last entries displayed in the table, one finds
$\Omega_{20}-\Omega_{19}\simeq1.96849-2.35043i \;$, which is close to the asymptotic spacing reported for intermediate-size SAdS black holes. This behavior is the finite-$x_+$ analogue of the nearly linear overtone structure that becomes more pronounced in the large-black-hole regime~\cite{CardosoKonoplyaLemos2003}.

Table~\ref{tables0ell0xp0_2xp100} extends the overtone comparison to two opposite regimes of the SAdS geometry. The case $x_+=0.2$ probes the small-black-hole regime, whereas $x_+=100$ is already deep in the large-black-hole regime. In both cases, the agreement with Ref. \cite{CardosoKonoplyaLemos2003} is excellent. For $x_+=100$, our spectral method reproduces all tabulated frequencies up to the displayed precision for $n\in\{0,\ldots,10\}$. For $x_+=0.2$, the agreement is likewise excellent for the fundamental mode and for the overtones $n\in\{2,\ldots,11\}$. The only visible exception is the imaginary part of the $n=1$ entry. Ref. \cite{CardosoKonoplyaLemos2003} quotes $\Omega=4.07086-0.98966i$, whereas our spectral result is $\Omega=4.070860-0.986629i$. This value is stable under changes of the spectral resolution, and the neighbouring overtones agree with the published data to the quoted accuracy. We therefore regard this as an isolated discrepancy in the published table, most likely attributable to tabulation or rounding, rather than as evidence of any deficiency in the spectral method.

The two halves of Table~\ref{tables0ell0xp0_2xp100} also illustrate the qualitative difference between small and large SAdS black holes. For $x_+=0.2$, the real parts are close to the pure-AdS scalar normal-mode sequence, while the imaginary parts remain comparatively small. This agrees with the expected small-black-hole behavior: as $x_+\to0$, the scalar QNM frequencies approach the pure-AdS normal modes $\Omega_{\rm AdS}=2n+\ell+3$, with the damping rate tending to zero. For $\ell=0$, the fundamental pure-AdS value is $\Omega=3$, and the $x_+=0.2$ fundamental mode, $\Omega=2.475112-0.389925i$, is already moving toward this limit. By contrast, for $x_+=100$, both the real and imaginary parts grow approximately linearly with the overtone number. This is consistent with the large-black-hole asymptotic behavior \cite{CardosoKonoplyaLemos2003}
\begin{equation}
\frac{\Omega_n}{x_+}\simeq(1.299-2.25i)n+1.856-2.673i \; ,\qquad n\to\infty \; ,
\end{equation}
or, equivalently,
\begin{equation}
\frac{\Omega_{n+1}-\Omega_n}{x_+}\simeq 1.299-2.25i \; .
\end{equation}
For example, using the last two modes displayed for $x_+=100$, we find
\begin{equation}
\frac{\Omega_{20}-\Omega_{19}}{100}\simeq 1.29914-2.25002i \; ,
\end{equation}
which is already extremely close to the predicted asymptotic spacing. This agreement also demonstrates that the spectral method remains accurate for a long overtone sequence.

As a final independent benchmark in the large-black-hole regime, Table~\ref{tab:daghigh_xp50} compares our scalar spectrum with the continued-fraction calculation of Ref. \cite{DaghighGreenMorey2023}. This comparison is useful because the continued-fraction approach is algorithmically independent of both the Horowitz--Hubeny power-series method and the present Chebyshev spectral discretization. The agreement is again at the level of the displayed digits for the low overtones, providing an additional validation of the implementation in the large-black-hole regime.

We next focus on the small-black-hole regime in Table~\ref{tab:smallBHKon2002}. In this limit, $x_+\ll1$, the SAdS scalar spectrum approaches the normal-mode spectrum of pure global AdS. The second column reports the fundamental $\ell=0$ scalar frequencies computed by Ref. \cite{Konoplya2002}, while the remaining columns show our spectral results for the fundamental mode and for the first three overtones. For the fundamental mode, the agreement with the data of Ref. \cite{Konoplya2002} is very good throughout the range of $x_+$. This provides a nontrivial check of the spectral method in a regime where the damping rates become small and the numerical problem is more delicate. The approach to the pure AdS limit is transparent in the same table. For a massless scalar field in four-dimensional global AdS, the normal-mode frequencies are $\Omega_{\rm AdS}=2n+\ell+3$. Thus, for $\ell=0$, the first four pure-AdS values are $3$, $5$, $7$, and $9$. As $x_+$ decreases, the real parts of our QNM frequencies move toward precisely these values. For example, at $x_+=1/100$ we find $\Omega_0\simeq2.97379-0.00055i$, $\Omega_1\simeq4.93781-0.00176i$, $\Omega_2\simeq6.89526-0.00382i$, and $\Omega_3\simeq8.84749-0.00694i$. These are small dissipative deformations of the pure-AdS values $3$, $5$, $7$, and $9$, confirming that the SAdS QNMs continuously approach the normal modes of empty AdS as the horizon radius tends to zero. The imaginary parts show the complementary trend: the damping decreases rapidly as $x_+$ is reduced. This is physically expected, since in the limit $x_+\to0$ the absorbing horizon shrinks away and the system approaches global AdS with reflecting boundary conditions. The small but nonzero negative imaginary parts in the last row therefore measure the weak absorption by a very small black hole. For the fundamental mode, this agrees with the conclusion in Ref. \cite{Konoplya2002} that $\Omega_I\to0^-$ as $x_+\to0$, while $\Omega_R$ tends to the corresponding AdS normal-mode frequency. The additional columns $n\in\{1,2,3\}$ extend the comparison beyond the fundamental mode tabulated in Ref. \cite{Konoplya2002}. They show that the same limiting behavior is not restricted to the fundamental frequency, since the first overtones also approach the expected AdS sequence $\Omega_{\rm AdS}=2n+3$ for $\ell=0$. There is one small isolated discrepancy in Table~\ref{tab:smallBHKon2002}, at $x_+=0.25$. Ref. \cite{Konoplya2002} quotes $\Omega_{\rm K}=2.41945-0.54735i$, whereas our spectral method gives $\Omega_{\rm SM}=2.41841-0.54799i$. The difference, $|\Omega_{\rm SM}-\Omega_{\rm K}|\simeq1.2\times10^{-3}$, is larger than a mere rounding effect. We repeated the computation and found the same value, stable under the changes of spectral resolution used in our convergence test. Since the neighboring entries $x_+=0.3$ and $x_+=0.2$ agree with the published data to the displayed accuracy, and since the $x_+=0.25$ value lies smoothly on the small-black-hole trend toward the pure-AdS normal mode, the comparison suggests an isolated discrepancy with the value quoted in Ref.~\cite{Konoplya2002}, rather than a numerical instability of the present method. We also stress that all present spectral-method frequencies quoted in the benchmark tables above satisfy the triplet-matching criterion described in Sec.~\ref{SecSpectralMethod}. Namely, each listed mode was identified independently at the three resolutions $N\in\{190,195,200\}$ and retained only if the largest pairwise difference between the matched frequencies satisfied the prescribed convergence criterion. Thus, the benchmark comparisons are not based on a single spectral resolution, but on modes that are stable under the resolution-triplet test. The quantity $\Delta\Omega_{\rm max}$, when quoted, should be understood as this resolution-spread diagnostic, rather than as a rigorous bound on the exact numerical error.

\begin{figure}[t]
\centering
\includegraphics[width=0.95\textwidth]{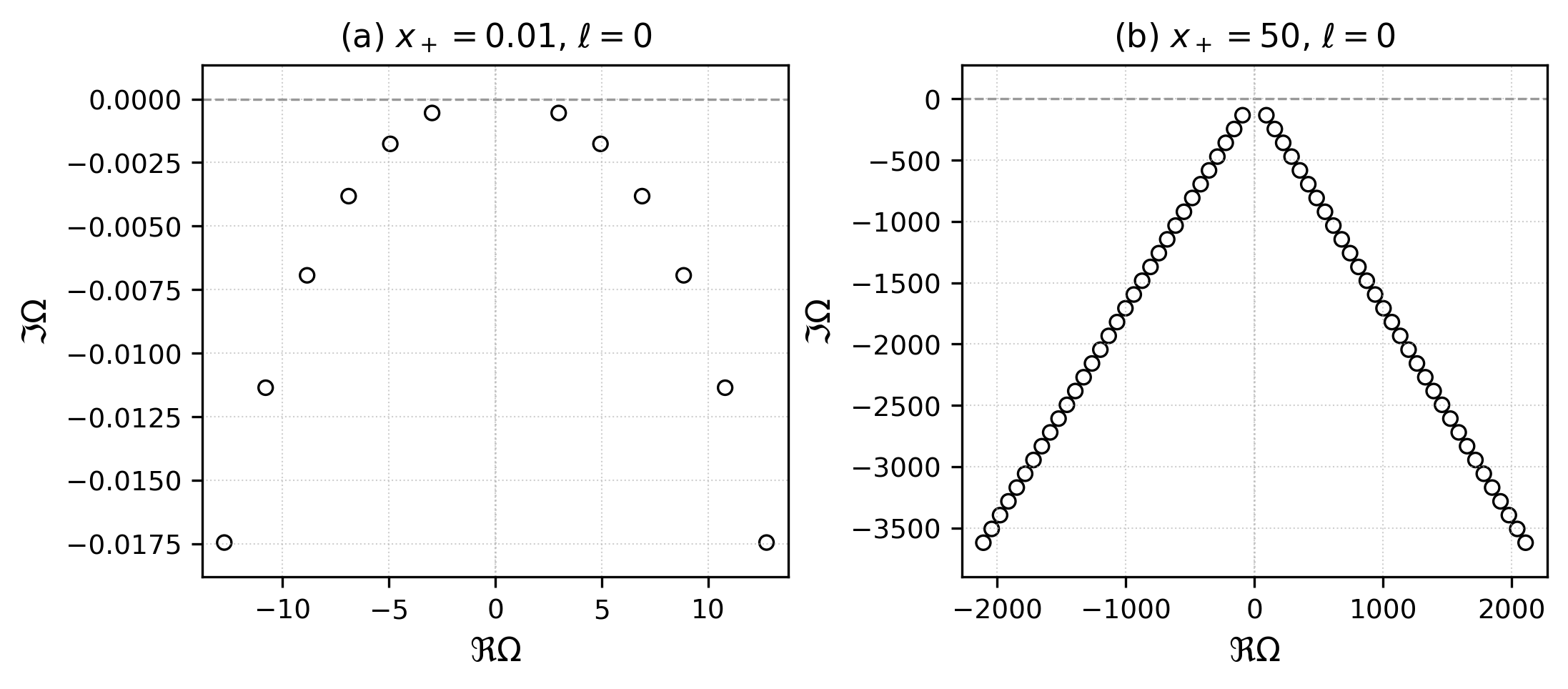}
\caption{\textit{
QNM spectrum of the massless scalar sector with $\ell=0$ and the standard vanishing-field boundary condition at the conformal AdS boundary. The left panel shows the small black hole case $x_+=0.01$, while the right panel shows the large black hole case $x_+=50$. For $x_+=0.01$, the modes lie very close to the real axis and approach the pure-AdS normal-mode values $\Omega_{\rm AdS}=2n+\ell+3$. For $x_+=50$, the modes organize into two approximately straight, symmetry-related branches, reflecting the nearly uniform high-overtone spacing of large SAdS black holes.
}}
\label{fig:qnm_scalar_xp001_xp50}
\end{figure}

Fig.~\ref{fig:qnm_scalar_xp001_xp50} provides a visual summary of the qualitative change between the small and large black-hole regimes. For $x_+=0.01$, the imaginary parts are very small and the real parts are close to the pure-AdS sequence $3,5,7,\ldots$, confirming the weakly damped normal-mode limit as $x_+\to0$. By contrast, for $x_+=50$ the modes are deeply damped and arrange themselves along two almost straight branches in the complex plane. The two branches are related by the symmetry $\Omega\mapsto-\Omega^*$, while the nearly constant spacing along each branch is the large-black-hole overtone behavior discussed above.

Finally, Table~\ref{tab:scalar_ell_scan} shows how the scalar spectrum changes with the angular number $\ell$ in three representative regimes of the SAdS geometry. We display the first five modes, $n\in\{0,\ldots,4\}$, for $x_+\in\{0.4,1,50\}$. The $x_+=0.4$ panel illustrates the transition toward the small-black-hole regime. As $\ell$ increases, the real part of the fundamental mode increases and moves toward the pure AdS value $\Omega_{\rm AdS}=\ell+3$, obtained from $\Omega_{\rm AdS}=2n+\ell+3$ at $n=0$. The values $\Omega_0(\ell=0)=2.36293-1.00647i$ and $\Omega_0(\ell=3)=5.18494-0.27708i$ are consistent with this trend. At the same time, the damping rate decreases as $\ell$ increases. This is expected because the angular-momentum barrier suppresses absorption by the black hole, making the modes longer lived. This behavior is the finite-$x_+$ version of the small-black-hole limit discussed in Ref. \cite{Konoplya2002}, where the imaginary part tends to zero and the real part approaches the normal-mode spectrum of pure AdS. For the intermediate case $x_+=1$, the same qualitative angular dependence is present, although the black-hole scale and the AdS scale are comparable. The fundamental mode shifts from $\Omega_0=2.79822-2.67121i$ at $\ell=0$ to $\Omega_0=5.00425-2.08948i$ at $\ell=3$. Increasing $\ell$ therefore raises the oscillation frequency and reduces the damping of the fundamental mode. For the overtones, however, the dependence on $\ell$ becomes progressively weaker, since the spacing between successive modes approaches an approximately uniform pattern, consistent with the overtone behavior already observed in the $\ell=0$ comparison tables. The large-black-hole panel, $x_+=50$, shows a qualitatively different feature. In this regime the spectrum is dominated by the horizon scale, and the dependence on $\ell$ is very weak. For example, the fundamental frequency changes only from $92.49368-133.19329i$ at $\ell=0$ to $92.59332-133.16520i$ at $\ell=3$, which is a small relative change compared with the overall scale of the frequency. The same weak angular dependence is visible for the first few overtones. This is consistent with the large SAdS-black-hole scaling $\Omega_n\propto x_+$. The large-black-hole data also display the expected nearly uniform overtone spacing. For instance, for $x_+=50$ and $\ell=0$, $\Omega_1-\Omega_0\simeq65.60708-112.63110i$, while $\Omega_4-\Omega_3\simeq65.01943-112.51774i$. Dividing by $x_+=50$, these spacings are already close to the asymptotic large-black-hole result $(\Omega_{n+1}-\Omega_n)/x_+\simeq1.299-2.25i$ reported in Ref. \cite{CardosoKonoplyaLemos2003}. Thus, the angular number mainly affects the low-lying offset of the spectrum, whereas the high-overtone spacing, especially for large black holes, is controlled primarily by $x_+$ and depends only weakly on $\ell$.

\subsection{The scalar sector with generalized boundary condition at the conformal AdS boundary}

We now turn to the generalized coefficient boundary condition $A\cos\theta+B\sin\theta=0$. The endpoint $\theta=0$ corresponds to the standard Dirichlet problem, while $\theta=\pi/2$ corresponds to the coefficient-Neumann condition. As emphasized above, this one-parameter family should be understood as a deformation of the minimally coupled massless scalar spectral problem, rather than as the standard alternative quantization available for scalars in the Breitenlohner-Freedman window. The main qualitative question is whether non-Dirichlet deformations preserve the stability of the scalar spectrum or generate additional modes with $\operatorname{Im}\Omega>0$.

\begin{figure*}[t]
\centering
\includegraphics[width=0.95\textwidth]{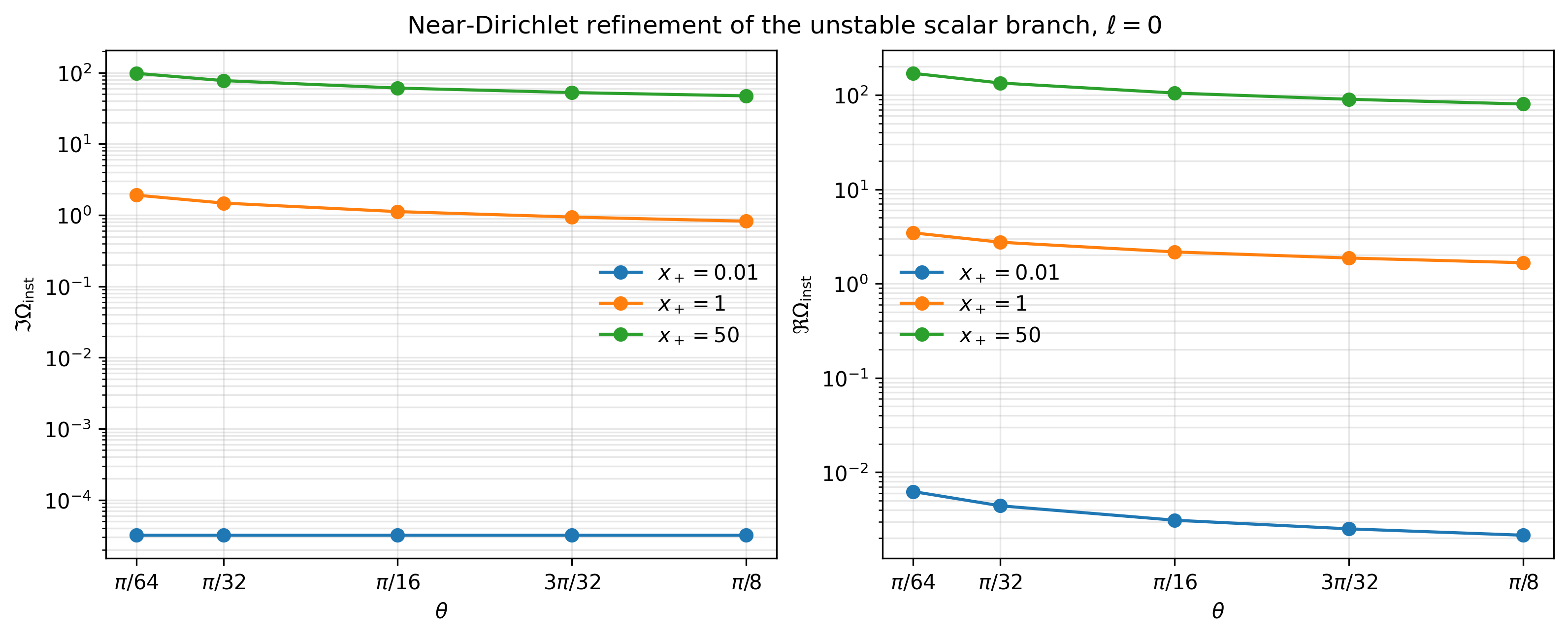}
\caption{\textit{
Near-Dirichlet refinement of the unstable scalar branch for the generalized
coefficient boundary condition $A\cos\theta+B\sin\theta=0$, with $\ell=0$. The left panel shows the growth rate $\operatorname{Im}\Omega_{\rm inst}$, while the right panel shows the oscillation frequency $\operatorname{Re}\Omega_{\rm inst}$. Both vertical axes are shown on a
logarithmic scale to display simultaneously the small, intermediate-, and
large-black hole regimes. The Dirichlet endpoint $\theta=0$ is not shown because no unstable mode is found there.
}}
\label{fig:near_dirichlet_instability}
\end{figure*}

Table~\ref{tab:generalized_instability} summarizes the unstable mode found for three representative black hole sizes, $x_+\in\{0.01,1,50\}$, and for several values of the boundary parameter $\theta$. The Dirichlet endpoint $\theta=0$ reproduces the standard scalar problem discussed in the previous subsection, and no unstable mode is found. By contrast, for every nonzero value of $\theta$ in the scan, an additional mode with $\operatorname{Im}\Omega>0$ appears. With the time dependence $e^{-i\omega t}$, such a mode grows exponentially in time. The unstable frequency is accompanied by its symmetry-related partner under $\Omega\mapsto-\Omega^*$. In the table we display only the branch with $\operatorname{Re}\Omega\geq0$. The instability is present in all three regimes probed here. The case $x_+=0.01$ is close to the global-AdS limit, $x_+=1$ represents an intermediate-size black hole, and $x_+=50$ lies deep in the large-black-hole regime. The appearance of the unstable branch in all three cases shows that, within the parameter values examined, the effect is not confined to a particular black-hole-size regime and is instead associated with the non-Dirichlet deformation of the massless scalar boundary condition. The growth rate is very small for $x_+=0.01$, of order $10^{-5}$, while it becomes of order unity for $x_+=1$ and of order $x_+$ for $x_+=50$. The small values of $\Delta\Omega_{\rm max}$ indicate that the listed
unstable modes are stable under the three-resolution matching procedure. For the intermediate and large black holes, the three spectral resolutions
often agree to extremely high precision. We do not interpret these very small
resolution-spreads as rigorous error bounds on the exact QNM frequencies.
They only indicate that the corresponding modes are very stable under the
resolution-triplet test. For this reason, whenever
$\Delta\Omega_{\rm max}<10^{-30}$, we quote the value simply as
$<10^{-30}$. The larger spreads in the $x_+=0.01$ case are consistent with
the greater numerical delicacy of the near-global-AdS regime, where the
unstable growth rates are very small and the modes lie close to the real axis.
\begin{figure*}[t]
\centering
\includegraphics[width=0.95\textwidth]{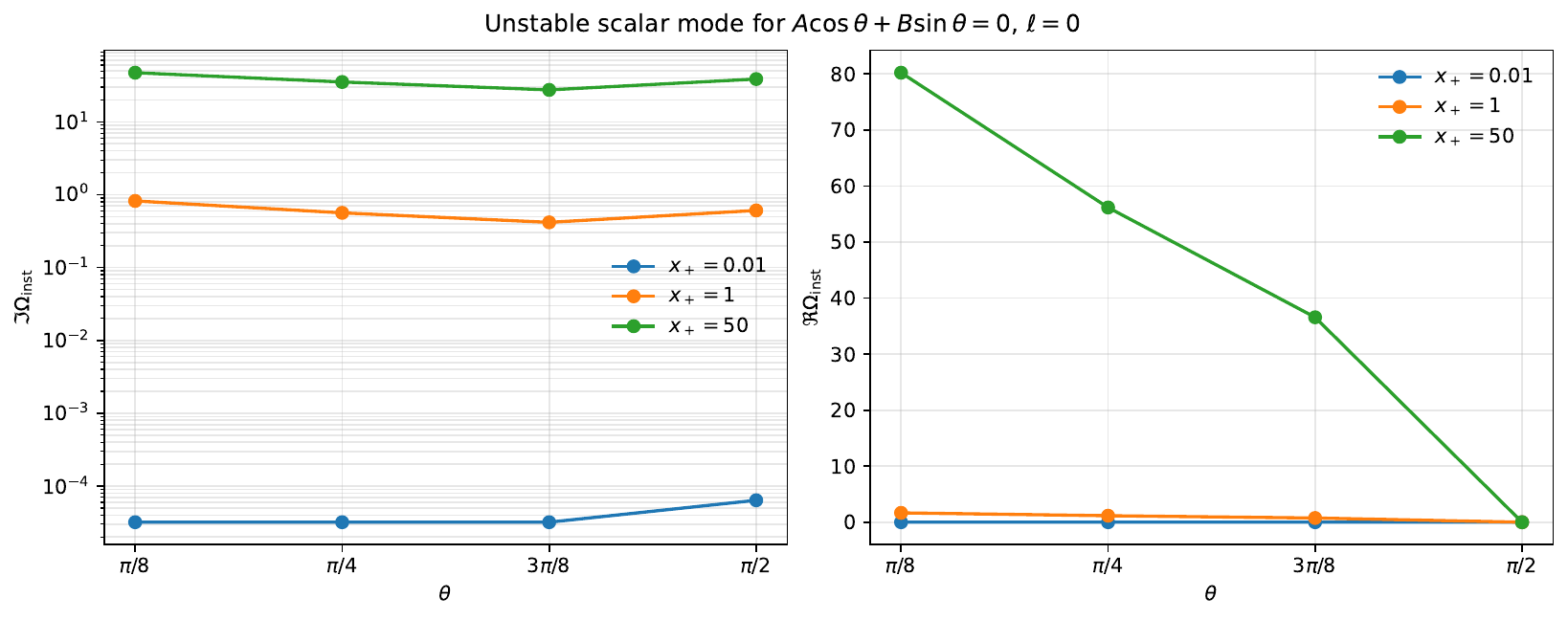}
\caption{\textit{
Unstable scalar mode for the generalized coefficient boundary condition $A\cos\theta+B\sin\theta=0$, with $\ell=0$. The left panel shows the growth
rate $\operatorname{Im}\Omega_{\rm inst}$, while the right panel shows the
oscillation frequency $\operatorname{Re}\Omega_{\rm inst}$, as functions of
$\theta$. The Dirichlet endpoint $\theta=0$ is not shown because no unstable
mode is found there. The growth-rate panel is shown on a logarithmic scale in
order to display simultaneously the small black hole, intermediate, and large-black hole regimes.
}}
\label{fig:generalized_instability_theta}
\end{figure*}
Fig.~\ref{fig:generalized_instability_theta} visualizes the dependence of the unstable branch on the boundary parameter. The growth rate remains positive for all nonzero values of $\theta$ shown, while the Dirichlet endpoint is stable. For $x_+=0.01$, the growth rate is very small, of order
$10^{-5}$, and is nearly independent of $\theta$ for $\theta\in\{\pi/8,\pi/4,3\pi/8\}$, before increasing at the coefficient-Neumann
endpoint. Although these growth rates would round to zero at four decimal
places, the unrounded values are strictly positive and are stable under the
three-resolution matching procedure. For $x_+=1$ and $x_+=50$, the same qualitative behavior is observed: the real part decreases as $\theta$ approaches $\pi/2$ and vanishes at the coefficient-Neumann endpoint, where the unstable mode becomes purely imaginary within our numerical classification tolerance, $|\Omega_R|<10^{-10}$. The large-black-hole case has much larger values of both $\operatorname{Re}\Omega_{\rm inst}$ and $\operatorname{Im}\Omega_{\rm inst}$, reflecting the dominant scale set by the horizon radius. To determine whether the instability appears only after a finite deformation away from the Dirichlet endpoint, we performed a near-Dirichlet refinement. In particular, we considered $\theta\in\{\pi/64,\pi/32,\pi/16,(3\pi)/32,\pi/8\}$, again for $x_+\in\{0.01,1,50\}$ and $\ell=0$. The resulting unstable branch is displayed in Fig.~\ref{fig:near_dirichlet_instability}, which 
shows that the unstable mode persists throughout the near-Dirichlet refinement. We find an unstable mode for every nonzero value of $\theta$ tested, down to $\theta=\pi/64$, and for all three representative black hole sizes. Thus, no finite critical angle is observed in the interval $\pi/64\leq\theta\leq\pi/8$. If a nonzero critical angle exists, it must satisfy $\theta_c<\pi/64$. The data are therefore consistent with the interpretation that the instability appears as soon as the Dirichlet condition is deformed, although the present scan cannot exclude a critical angle below $\pi/64$. The behavior close to $\theta=0$ depends on the black hole size. For the small black hole, $x_+=0.01$, the growth rate remains approximately constant, $\operatorname{Im}\Omega_{\rm inst}\simeq3.19\times10^{-5}$, throughout the near-Dirichlet range shown. For $x_+=1$ and $x_+=50$, both $\operatorname{Re}\Omega_{\rm inst}$ and
$\operatorname{Im}\Omega_{\rm inst}$ increase as $\theta$ is decreased. The numerical data suggest that the unstable mode is not simply one of the ordinary damped QNMs continuously crossing the real axis as $\theta$ is varied. In the near-Dirichlet refinement, the unstable branch persists for the
smallest nonzero angles considered, while the Dirichlet endpoint itself contains no corresponding unstable mode. This points to the possibility that the unstable branch is a separate boundary-condition-induced branch of the spectrum. A definitive characterization of its analytic origin, and in particular of its limiting behaviour as $\theta\to 0^+$, remains an open question. These results demonstrate that, within the generalized coefficient problem considered here, the standard Dirichlet boundary condition is not merely a technical choice for the minimally coupled massless scalar. Deforming the relation between the slow and fast asymptotic coefficients changes the stability properties of the spectrum. The damped QNM tower remains present, but the new and physically most relevant feature is the appearance of an unstable mode for every nonzero value of \(\theta\) considered in our scan.

\begin{table}[t]
\centering
\caption{\textit{Fundamental scalar modes across different SAdS black hole sizes $x_+$. We compare our spectral method results with $N=200$ Chebyshev polynomials (final column) against the Horowitz--Hubeny power series method of \cite{HorowitzPRD2000} (second column), and the matrix method of \cite{LinCQG2017} (third column) based on interpolation/discretization of the master equation. SM stands for spectral method.}}
\label{tables0ell0n0}
\setlength\tabcolsep{0.1cm}
\def\arraystretch{1.5}
 \begin{tabular}{@{}|c| c| c| c|@{}} 
 \hline 
 $x_+$  & $\Omega_{\rm HH}$ \cite{HorowitzPRD2000} & $\Omega_{\rm LIN}$ \cite{LinCQG2017} & $\Omega_{\rm SM}$\\ [0.5ex]
 \hline
$0.4$   & $2.3629-1.0064i$     & $2.38152-0.938149i$ & $2.3629-1.0065i$ \\ 
        \hline
$0.6$   & $2.4316-1.5797i$     & $2.4316-1.5797i$     & $2.4316-1.5797i$ \\ 
        \hline
$0.8$   & $2.5878-2.1304i$     & $2.5878-2.1304i$     & $2.5878-2.1304i$ \\ 
        \hline
$1$     & $2.7982-2.6712i$     & $2.7982-2.6712i$     & $2.7982-2.6712i$ \\ 
        \hline
$5$     & $9.4711-13.3255i$    & $9.4711-13.3255i$    & $9.4711-13.3255i$ \\ 
        \hline
$10$    & $18.6070-26.6418i$   & $18.6070-26.6418i$   & $18.6070-26.6418i$ \\ 
        \hline
$50$    & $92.4937-133.1933i$  & $92.4937-133.1933i$  & $92.4937-133.1933i$ \\ 
        \hline
$100$   & $184.9534-266.3856i$ & $184.9534-266.3856i$ & $184.9534-266.3856i$ \\ [0.5ex]
\hline
\end{tabular}
\end{table}
\begin{table}[t]
\centering
\caption{\textit{Detailed overtone comparison at $x_+=1$ and $\ell=0$. We compare our
spectral method results, obtained with $N=200$ Chebyshev polynomials and
checked against $N=190,195$, with the data obtained by means of the Horowitz--Hubeny power series method \cite{CardosoKonoplyaLemos2003} and the asymptotic iteration
method of \cite{ChoCornellDoukasNaylor2010} with $70$ iterations. Entries labelled N/A
indicate unavailable data. SM stands for spectral method. $\dagger$: The value $8.46153-9.74852i$ appearing in \cite{CardosoKonoplyaLemos2003} for this mode is a known mismatch in the real part and it was already reported by \cite{ChoCornellDoukasNaylor2010}.}}
\label{tables0ell0xp1}
\setlength\tabcolsep{0.1cm}
\def\arraystretch{1.5}
 \begin{tabular}{@{}|c| c| c| c|@{}} 
 \hline
$n$  & $\Omega_{\rm CKL}$ \cite{CardosoKonoplyaLemos2003} & $\Omega_{\rm AIM}$ \cite{ChoCornellDoukasNaylor2010} & $\Omega_{\rm SM}$\\ [0.5ex]
 \hline
$0$   & $2.7982-2.6712i$       & $2.79823-2.67121i$ & $2.7982-2.6712i$ \\ 
        \hline
$1$   & $4.75849-5.03757i$     & $4.75850-5.03757i$ & $4.7585-5.0376i$ \\ 
        \hline
$2$   & $6.71927-7.39449i$     & $6.71931-7.39450i$ & $6.7193-7.3945i$ \\ 
        \hline
$3$   & $8.46153^\dagger-9.74852i$     & $8.68233-9.74854i$ & $8.6822-9.7485i$ \\ 
        \hline
$4$   & $10.6467-12.1012i$     & $10.6469-12.1013i$ & $10.6467-12.1012i$ \\ 
        \hline
$5$   & $12.6121-14.4533i$     & $12.6125-14.4533i$ & $12.6121-14.4533i$ \\ 
        \hline
$6$   & $14.5782-16.8049i$     & $14.5788-16.8050i$ & $14.5782-16.8049i$ \\ 
        \hline
$7$   & $16.5449-19.1562i$     & $16.5457-19.1563i$ & $16.5449-19.1562i$ \\ 
        \hline
$8$   & $18.5119-21.5073i$     & \mbox{N/A}         & $18.5119-21.5073i$ \\ 
        \hline
$9$   & $20.4792-23.8583i$     & \mbox{N/A}         & $20.4792-23.8583i$ \\ 
        \hline
$10$  & $22.44671-26.20913i$   & \mbox{N/A}         & $22.4467-26.2091i$ \\ 
        \hline
$11$  & $24.41443-28.55989i$   & \mbox{N/A}         & $24.4144-28.5599i$ \\ 
        \hline
$12$  & $26.38230-30.91059i$   & \mbox{N/A}         & $26.3823-30.9106i$ \\   
        \hline
$13$  & $28.35029-33.26123i$   & \mbox{N/A}         & $28.3503-33.2612i$ \\ 
        \hline
$14$  & $30.31839-35.61183i$   & \mbox{N/A}         & $30.3184-35.6118i$ \\ 
        \hline
$15$  & $32.28658-37.96238i$   & \mbox{N/A}         & $32.2866-37.9624i$ \\ 
        \hline
$16$  & $34.25485-40.31290i$   & \mbox{N/A}         & $34.2548-40.3129i$ \\ 
        \hline
$17$  & $36.22318-42.66340i$   & \mbox{N/A}         & $36.2232-42.6634i$ \\ 
        \hline
$18$  & $38.19157-45.01387i$   & \mbox{N/A}         & $38.1916-45.0139i$ \\ 
       \hline
$19$  & $40.16002-47.36431i$   & \mbox{N/A}         & $40.1600-47.3643i$ \\ 
       \hline
$20$  & \mbox{N/A}             & \mbox{N/A}         & $42.1285-49.7147i$ \\ [0.5ex]
\hline
\end{tabular}
\end{table}

\begin{table}[t]
\centering
\caption{\textit{Detailed overtone comparison at $x_+\in\{0.2,100\}$ and $\ell=0$. We compare our
spectral method results, obtained with $N=200$ Chebyshev polynomials and
checked against $N=190,195$, with the data obtained by means of the Horowitz-Hubeny power series method \cite{CardosoKonoplyaLemos2003}. Entries labelled N/A
indicate unavailable data. SM stands for spectral method.}}
\label{tables0ell0xp0_2xp100}
\setlength\tabcolsep{0.1cm}
\def\arraystretch{1.5}
 \begin{tabular}{@{}|c| c| c| c| c| c| c| c|@{}} 
 \hline
$x_+$ & $n$  & $\Omega_{\rm CKL}$ \cite{CardosoKonoplyaLemos2003} &  $\Omega_{\rm SM}$ & $x_+$ & $n$  & $\Omega_{\rm CKL}$ \cite{CardosoKonoplyaLemos2003} &  $\Omega_{\rm SM}$\\ [0.5ex]
 \hline
$0.2$ & $0$     & $2.47511-0.38990i$  & $2.4751-0.3899i$ & $100$ & $0$ & $184.95344-266.38560i$  & $184.9534-266.3856i$ \\ 
      \hline
      & $1$     & $4.07086-0.98966i$  & $4.0709-0.9866i$  &      & $1$ & $ 316.14466-491.64354i$ & $316.1447-491.6435i$  \\ 
        \hline
      &$2$      & $5.72783-1.57600i$  & $5.7278-1.5760i$  &      & $2$ & $446.46153-716.75722i$  & $446.4615-716.7572i$  \\  
        \hline
      &$3$      & $7.40091-2.15869i$  & $7.4009-2.1587i$  &      & $3$ & $576.55983-941.81253i$  & $576.5598-941.8125i$  \\  
        \hline
      &$4$      & $9.08118 -2.73809i$ & $9.0812-2.7381i$  &      & $4$ & $706.57518-1166.8440i$  & $706.5752-1166.8441i$  \\  
        \hline
      &$5$      & $10.7655-3.31557i$  & $10.7655-3.3156i$ &      & $5$ & $836.55136-1391.8641i$  & $836.5514-1391.8641i$  \\  
        \hline
      &$6$      & $12.45222-3.89179i$ & $12.4522-3.8918i$ &     & $6$ & $966.50635-1616.8779i$  & $966.5063-1616.8780i$  \\  
        \hline
      &$7$      & $14.14065-4.46714i$ & $14.1406-4.4671i$ &      & $7$ & $1096.44876-1841.88813i$ & $1096.4488-1841.8881i$  \\  
        \hline
      &$8$      & $15.83026-5.04186i$ & $15.8303-5.0419i$ &      & $8$ & $1226.38317-2066.89596i$ & $1226.3832-2066.8960i$  \\  
        \hline
      &$9$      & $17.52070-5.61610i$ & $17.5207-5.6161i$ &      & $9$ & $1356.31222 -2291.90222i$ & $1356.3122-2291.9022i$  \\  
        \hline
      &$10$     & $19.21191-6.18997i$ & $19.2119-6.1900i$ &      & $10$& $1486.23753 -2516.90740i$ & $1486.2375-2516.9074i$  \\ 
        \hline
      &$11$     & $ 20.90359-6.76355i$& $20.9036-6.7635i$ &      & $11$ & \mbox{N/A}               & $1616.1601-2741.9118i$  \\  
        \hline
      &$12$     & \mbox{N/A}          & $22.5957-7.3369i$ &      & $12$ & \mbox{N/A}               & $1746.0808-2966.9156i$  \\     
        \hline
      &$13$     & \mbox{N/A}          & $24.2881-7.9100i$ &      & $13$ & \mbox{N/A}               & $1875.9998-3191.9190i$  \\   
        \hline
      &$14$     & \mbox{N/A}          & $25.9809-8.4830i$ &      & $14$ & \mbox{N/A}               & $2005.9177-3416.9220i$  \\   
        \hline
      &$15$     & \mbox{N/A}          & $27.6738-9.0558i$ &      & $15$ & \mbox{N/A}               & $2135.8347-3641.9248i$  \\ 
        \hline
      &$16$     & \mbox{N/A}          & $29.3670-9.6285i$ &      & $16$ & \mbox{N/A}               & $2265.7509-3866.9274i$  \\ 
        \hline
      &$17$     & \mbox{N/A}          & $31.0603-10.2011i$ &     & $17$ & \mbox{N/A}               & $2395.6665-4091.9298i$  \\  
        \hline
      &$18$     & \mbox{N/A}          & $32.7538-10.7737i$ &     & $18$ & \mbox{N/A}               & $2525.5817-4316.9320i$  \\   
       \hline
      &$19$     & \mbox{N/A}          & $34.4474-11.3461i$ &     & $19$ & \mbox{N/A}               & $2655.4964-4541.9342i$  \\   
       \hline
      &$20$     & \mbox{N/A}          & $36.1412-11.9185i$ &     & $20$ & \mbox{N/A}               & $2785.4107-4766.9362i$  \\ [0.5ex]
\hline
\end{tabular}
\end{table}

\begin{table}[t]
\centering
\caption{\textit{
Independent continued-fraction benchmark for the scalar sector at $x_+=50$ and $\ell=0$. We compare our spectral method frequencies obtained with $N=200$ Chebyshev polynomials and checked against $N=190,195$ with the continued-fraction results of \cite{DaghighGreenMorey2023}. Their overtone label starts at $n=1$. Here we relabel the fundamental mode as $n=0$. SM stands for spectral method.
}}
\label{tab:daghigh_xp50}
\setlength\tabcolsep{0.15cm}
\def\arraystretch{1.35}
\begin{tabular}{@{}|c|c|c|@{}}
\hline
$n$ & $\Omega_{\rm DGM}$ \cite{DaghighGreenMorey2023} & $\Omega_{\rm SM}$ \\
\hline
0 & $92.4936786-133.1932905i$  & $92.4937-133.1933i$ \\
\hline
1 & $158.1007632-245.8243897i$ & $158.1008-245.8244i$ \\
\hline
2 & $223.2708309-358.3832592i$ & $223.2708-358.3833i$ \\
\hline
3 & $288.3316967-470.9129117i$ & $288.3317-470.9129i$ \\
\hline
4 & $353.3511287-583.4306502i$ & $353.3511-583.4307i$ \\
\hline
5 & $418.3509995-695.9426546i$ & $418.3510-695.9427i$ \\
\hline
\end{tabular}
\end{table}

\begin{table}[t]
\centering
\caption{\textit{Small-black-hole scalar spectrum for $\ell=0$ and several values of
$x_+$. We compare the fundamental mode obtained with our spectral method
with the small-black-hole data of \cite{Konoplya2002}. The latter
were obtained using a Horowitz--Hubeny-type power-series expansion adapted to
the small-$x_+$ regime. Our results were computed with $N=200$ Chebyshev
polynomials and checked against $N=190,195$. Entries labelled N/A indicate
unavailable data, while SM stands for spectral method.}}
\label{tab:smallBHKon2002}
\setlength\tabcolsep{0.15cm}
\def\arraystretch{1.35}
\begin{tabular}{@{}|c|c|c|c|c|c|@{}}
\hline
$x_+$ & $\Omega_{\rm K}$ $(n=0)$ \cite{Konoplya2002} & $\Omega_{\rm SM}$ $(n=0)$ & $\Omega_{\rm SM}$ $(n=1)$ & $\Omega_{\rm SM}$ $(n=2)$ & $\Omega_{\rm SM}$ $(n=3)$\\
\hline
$0.3$  & $2.38447-0.70413i$  & $2.3845-0.7041i$ & $3.9836-1.5467i$ & $5.6193-2.3692i$ & $7.2658-3.1857i$\\
\hline
$0.25$ & $2.41945-0.54735i$  & $2.4184-0.5480i$ & $4.0141-1.2726i$ & $5.6571-1.9800i$ & $7.3131-2.6811i$\\
\hline
$0.2$  & $2.47511-0.3899i$   & $2.4751-0.3899i$ & $4.0709-0.9866i$ & $5.7278-1.5760i$ & $7.4009-2.1587i$\\
\hline
$0.125$& $2.62274-0.16392i$  & $2.6227-0.1639i$ & $4.2386-0.5247i$ & $5.9267-0.9242i$ & $7.6426-1.3201i$ \\
\hline
$0.1$  & $2.69282-0.10096i$  & $2.6928-0.1010i$ & $4.3352-0.3626i$ & $6.0351-0.6876i$ & $7.7695-1.0168i$ \\
\hline
$1/12$ & $2.74472-0.06616i$  & $2.7447-0.0662i$ & $4.4189-0.2567i$ & $6.1288-0.5232i$ & $7.8759-0.8045i$ \\
\hline
$1/14$ & $2.78341-0.04578i$  & $2.7835-0.0458i$ & $4.4902-0.1857i$ & $6.2114-0.4037i$ & $7.9674-0.6470i$ \\
\hline
$1/16$ & $2.81289-0.03311i$  & $2.8129-0.0331i$ & $4.5501-0.1373i$ & $6.2847-0.3147i$ & $8.0480-0.5258i$ \\
\hline
$1/18$ & $2.83574-0.02491i$  & $2.8357-0.0249i$ & $4.6001-0.1038i$ & $6.3498-0.2477i$ & $8.1202-0.4305i$ \\
\hline
$1/20$ & $2.8539-0.01932i$   & $2.8539-0.0193i$ & $4.6418-0.0802i$ & $6.4075-0.1969i$ & $8.1854-0.3544i$ \\
\hline
$1/25$ & $2.88584-0.01138i$  & $2.8859-0.0114i$ & $4.7191-0.0458i$ & $6.5233-0.1161i$ & $8.3233-0.2231i$ \\
\hline
$1/30$ & $2.9065-0.0074i$    & $2.9065-0.0075i$ & $4.7707-0.0289i$ & $6.6071-0.0732i$ & $8.4311-0.1455i$ \\
\hline
$1/100$& N/A                 & $2.9738-0.0006i$ & $4.9378-0.0018i$ & $6.8953-0.0038i$ & $8.8475-0.0069i$ \\
\hline
\end{tabular}
\end{table}

\begin{table*}[t]
\centering
\caption{\textit{
First five scalar modes for selected values of the black hole size $x_+$
and angular number $\ell$. We use the standard vanishing-field boundary
condition at the AdS boundary. The frequencies were obtained with the spectral
method using $N=200$ Chebyshev polynomials and checked against
$N=190,195$.}}
\label{tab:scalar_ell_scan}
\setlength\tabcolsep{0.08cm}
\def\arraystretch{1.35}
\scriptsize
\begin{tabular}{@{}|c|c|c|c|c|c|@{}}
\hline
\multicolumn{6}{|c|}{$x_+=0.4$}\\
\hline
$\ell$ & $\Omega_0$ & $\Omega_1$ & $\Omega_2$ & $\Omega_3$ & $\Omega_4$\\
\hline
$0$ & $2.3629-1.0065i$ & $3.9786-2.0728i$ & $5.6171-3.1196i$  & $7.2637-4.1611i$ & $8.9145-5.2002i$ \\
\hline
$1$ & $3.1338-0.7029i$ & $4.5679-1.7813i$ & $6.1146-2.8525i$  & $7.7035-3.9152i$ & $9.3136-4.9717i$ \\
\hline
$2$ & $4.1285-0.4512i$ & $5.3977-1.4466i$ & $6.8358-2.5038i$  & $8.3492-3.5712i$ & $9.9042-4.6378i$ \\
\hline
$3$ & $5.1849-0.2771i$ & $6.3383-1.1521i$ & $7.6794-2.1645i$  & $9.1155-3.2154i$ & $10.6096-4.2786i$ \\
\hline

\multicolumn{6}{|c|}{$x_+=1$}\\
\hline
$\ell$ & $\Omega_0$ & $\Omega_1$ & $\Omega_2$ & $\Omega_3$ & $\Omega_4$\\
\hline
$0$ & $2.7982-2.6712i$  & $4.7585-5.0376i$ & $6.7193-7.3945i$ & $8.6822-9.7485i$ & $10.6467-12.1013i$  \\
\hline
$1$ & $3.3306-2.4895i$  & $5.1724-4.8881i$ & $7.0710-7.2646i$ & $8.9940-9.6321i$ & $10.9298-11.9948i$ \\
\hline
$2$ & $4.1171-2.2756i$  & $5.8165-4.6678i$ & $7.6311-7.0561i$ & $9.4978-9.4367i$ & $11.3927-11.8112i$ \\
\hline
$3$ & $5.0043-2.0895i$  & $6.5791-4.4385i$ & $8.3067-6.8206i$ & $10.1116-9.2058i$& $11.9603-11.5878i$ \\
\hline

\multicolumn{6}{|c|}{$x_+=50$}\\
\hline
$\ell$ & $\Omega_0$ & $\Omega_1$ & $\Omega_2$ & $\Omega_3$ & $\Omega_4$\\
\hline
$0$ & $92.4937-133.1933i$ & $158.1008-245.8244i$ & $223.2708-358.3833i$ & $288.3317-470.9129i$ & $353.3511-583.4307i$ \\
\hline
$1$ & $92.5103-133.1886i$ & $158.1134-245.8211i$ & $223.2814-358.3805i$& $288.3410-470.9105i$ & $353.3595-583.4285i$ \\
\hline
$2$ & $92.5435-133.1792i$ & $158.1387-245.8144i$ & $223.3026-358.3749i$ & $288.3595-470.9056i$ & $353.3762-583.4241i$ \\
\hline
$3$ & $92.5933-133.1652i$ & $158.1765-245.8044i$ & $223.3343-358.3666i$ & $288.3873-470.8983i$ & $353.4013-583.4175i$ \\
\hline
\end{tabular}
\end{table*}

\begin{table*}[t]
\centering
\caption{\textit{Unstable scalar mode for the generalized coefficient boundary condition
$A\cos\theta+B\sin\theta=0$, with $\ell=0$. We display only the branch
with $\Omega_R\geq0 $. The partner with negative real part is obtained from
$\Omega\mapsto-\Omega^*$. The entry ``none'' at $\theta=0$ corresponds to
the standard Dirichlet problem, for which no unstable mode is found. The
frequencies were obtained with the spectral method using $N=200$ Chebyshev
polynomials and checked against $N\in\{90,195\}$. Here
$\Delta\Omega_{\rm max}$ denotes the resolution-spread of the triplet
matching procedure, namely the largest pairwise difference between the matched
frequencies obtained at $N\in\{190,195,200\}$. It is used as an internal
convergence diagnostic rather than as a rigorous error bound. When the resolution-spread is smaller than $10^{-30}$, we report it simply as $<10^{-30}$. Entries of the form $|\Omega_R|<10^{-10}$ denote modes classified as purely imaginary within
the numerical tolerance used in the mode classification.
}}
\label{tab:generalized_instability}
\setlength\tabcolsep{0.12cm}
\def\arraystretch{1.3}
\begin{tabular}{@{}|c|c|c|c|c|@{}}
\hline
\(x_+\) & \(\theta\) & \(\Omega_R^{\rm inst}\) &
\(\Omega_I^{\rm inst}\) & \(\Delta\Omega_{\rm max}\) \\
\hline
\(0.01\) & \(0\)      & none & none & -- \\
\(0.01\) & \(\pi/8\)  & \(2.1505\times10^{-3}\) & \(3.1934\times10^{-5}\) & \(2.76\times10^{-11}\)  \\
\(0.01\) & \(\pi/4\)  & \(1.3838\times10^{-3}\) & \(3.1934\times10^{-5}\) & \(1.78\times10^{-11}\) \\
\(0.01\) & \(3\pi/8\) & \(8.9030\times10^{-4}\) & \(3.1934\times10^{-5}\) & \(1.15\times10^{-11}\) \\
\(0.01\) & \(\pi/2\)  & \(|\Omega_R|<10^{-10}\) & \(6.3868\times10^{-5}\) & \(3.12\times10^{-12}\) \\
\hline
\(1\) & \(0\)         & none & none & -- \\
\(1\) & \(\pi/8\)     & \(1.6658\) & \(0.8213\) & \(<10^{-30}\) \\
\(1\) & \(\pi/4\)     & \(1.1590\) & \(0.5634\) & \(<10^{-30}\) \\
\(1\) & \(3\pi/8\)    & \(0.7483\) & \(0.4174\) & \(<10^{-30}\) \\
\(1\) & \(\pi/2\)     & \(|\Omega_R|<10^{-10}\) & \(0.6081\) & \(<10^{-30}\)\\
\hline
\(50\) & \(0\)        & none & none & -- \\
\(50\) & \(\pi/8\)    & \(80.2260\) & \(47.2803\) & \(<10^{-30}\) \\
\(50\) & \(\pi/4\)    & \(56.1688\) & \(35.2695\) & \(<10^{-30}\) \\
\(50\) & \(3\pi/8\)   & \(36.5749\) & \(27.4649\) & \(<10^{-30}\) \\
\(50\) & \(\pi/2\)    & \(|\Omega_R|<10^{-10}\) & \(38.6960\) & \(<10^{-30}\) \\
\hline
\end{tabular}
\end{table*}


\section{Conclusions and outlook \label{sec:conclusions}}

In this work, we have studied QNMs of a minimally coupled massless scalar field on four-dimensional SAdS-black hole. The analysis was organized around two related spectral problems. The first is the standard scalar SAdS problem, defined by purely ingoing behavior at the event horizon and the Dirichlet, or vanishing-field, condition at the conformal AdS boundary. The second is a generalized coefficient problem in which the two independent asymptotic data of the same massless scalar are related by a one-parameter boundary condition interpolating between the Dirichlet endpoint and a coefficient-Neumann endpoint.

On the numerical side, we developed a Chebyshev spectral implementation in which the radial equation is compactified to a finite interval and recast as a quadratic matrix pencil in the dimensionless frequency $\Omega=\omega R$. For the Dirichlet problem, both the ingoing horizon behavior and the AdS-boundary falloff are incorporated directly into the radial ansatz. For the generalized coefficient problem, the ingoing horizon behavior is factored out, while the boundary condition at the conformal AdS boundary is imposed as a spectral boundary row. In the latter case, the boundary row makes the leading quadratic matrix singular, and we therefore implemented the problem by eliminating the boundary constraint before solving the reduced quadratic pencil. In both cases, physical modes were extracted by matching spectra obtained at the three resolutions $N\in\{190,195,200\}$, and only modes stable under this triplet procedure were retained.

For the standard Dirichlet scalar problem, the method reproduces the known SAdS scalar spectrum with high accuracy. We compared against the Horowitz-Hubeny power-series method, Konoplya's small-black-hole data, the Cardoso-Konoplya-Lemos overtone tables, the asymptotic iteration method, the Lin-Qian matrix method, and the continued-fraction results of Daghigh, Green and Morey. The agreement across these independent methods provides a stringent validation of the spectral construction. The calculations also reproduce the expected limiting behaviors. In the small-black-hole regime, the scalar frequencies approach the pure AdS normal-mode sequence and the damping rate tends to zero. In the large-black-hole regime, the frequencies scale linearly with the horizon radius and the highly damped modes approach the known nearly uniform spacing in the complex-frequency plane. The angular-momentum scan further confirms that $\ell$ strongly affects the low-lying modes for small and intermediate black holes, while its relative effect becomes weak in the large-black-hole regime, where the dominant scale is set by $x_+=r_+/R$.

The generalized coefficient boundary condition leads to a qualitatively different result. The Dirichlet endpoint reproduces the stable standard spectrum. However, for every nonzero value of the boundary parameter examined, we find an additional mode with positive imaginary part. With the convention $e^{-i\omega t}$, this corresponds to exponential growth in time. The unstable mode appears together with its symmetry-related partner under $\Omega\mapsto-\Omega^*$, and it is present for small, intermediate, and large black holes. A near-Dirichlet refinement shows that the unstable branch persists down to the smallest deformation we probed, $\theta=\pi/64$. Thus, no finite critical angle was observed in our numerical scan. We do not claim this as a proof that the instability exists for every arbitrarily small nonzero deformation, but the data are consistent with the interpretation that the Dirichlet endpoint is a special stable point of the generalized coefficient
problem. We emphasize again that the generalized coefficient boundary condition studied here is not the standard alternative quantization of a scalar field in the Breitenlohner--Freedman window. It is a controlled deformation of the minimally coupled massless scalar QNM problem, formulated by relating the two independent asymptotic coefficients of that field. The instability found here should therefore be interpreted as a statement about this deformed massless scalar spectral problem. In particular, it shows that the stability of the standard Dirichlet spectrum is not automatically inherited by arbitrary non-Dirichlet coefficient relations at the conformal AdS boundary.

Several extensions are natural. The first is to understand analytically the origin of the unstable branch. In particular, it would be useful to determine whether the branch can be followed perturbatively away from the Dirichlet endpoint, whether its frequency diverges or leaves the physical sheet as $\theta\to 0$, and whether one can establish a rigorous instability criterion for the generalized coefficient boundary condition. A complementary time-domain study would also be valuable. The frequency-domain calculation identifies exponentially growing modes, but direct evolution would show how these modes dominate generic initial data and would clarify the role of the boundary condition in the associated energy balance.

A second direction is to repeat the analysis for scalar fields whose two AdS falloffs are both admissible in the standard sense. In particular, conformally coupled or massive scalars in the appropriate mass range admit genuine Robin or mixed boundary conditions related to the usual alternative quantization. Studying those cases with the same spectral machinery would allow a clean comparison between the generalized coefficient deformation considered here and physically admissible Robin families in the Ishibashi--Wald framework. It would also make contact with recent work on Robin boundary conditions and boundary-condition-induced instabilities in AdS black-hole backgrounds.

A third and immediate extension is to apply the present spectral method to the electromagnetic and gravitational sectors. This is particularly important because boundary conditions for higher-spin perturbations in AdS are known to be more subtle than simple field vanishing. For Maxwell perturbations, vanishing energy flux at the AdS boundary permits more than one reflecting branch, and the second branch gives spectra distinct from the older Dirichlet-type calculations. For gravitational perturbations, the axial and polar sectors require special care because fixed-boundary-metric or AdS/CFT-motivated prescriptions lead naturally to parity-dependent boundary conditions, with Dirichlet-type behavior in the axial sector and Robin-type behavior in the polar sector. The matrix-pencil formulation developed here is well suited to these problems, since Dirichlet, Neumann, Robin, and more general coefficient conditions can all be incorporated at the level of spectral boundary rows.

Beyond spherical SAdS, it would also be interesting to consider charged, rotating, or topological AdS black holes. In those settings, one expects a richer interplay between horizon physics, boundary conditions, and possible instabilities. Another promising direction is to use the high-precision spectra obtained here to investigate non-normality and pseudospectral sensitivity of AdS QNM problems. Since the QNM operator is non-self-adjoint, small changes in the boundary condition or in the operator coefficients can reorganize the spectrum in ways not captured by eigenvalues alone. The generalized boundary-condition-induced instability found in this work provides a concrete setting in which such questions can be explored.


\subsection*{Code availability}

\noindent The codes used to assemble the spectral matrices and to post-process the quadratic eigenvalue problems are available at the following URL address:
\begin{center}
  \url{https://github.com/dutykh/schwarzschild-ads/}
\end{center}


\section*{Acknowledgements}
ASC was partly supported by the National Research Foundation of South Africa.


\bibliographystyle{aipnum4-2}
\bibliography{QNMS_v2}

\end{document}